\documentclass[]{aa}  

\usepackage{graphicx}
\usepackage{txfonts}
\usepackage{siunitx}
\usepackage[colorlinks]{hyperref}
\usepackage[normalem]{ulem}
\newcommand{\rsun}{R$_{\sun}$}
\newcommand{\kms}{km\,s$^{-1}$}
\newcommand{\Lya}{Ly-$\alpha$}

\begin{document} 

\title{A Coronal Mass Ejection followed by a prominence eruption \\
and a plasma blob as observed by Solar Orbiter}

\author{
  A.~Bemporad\inst{\ref{INAF-OATo}}
  \and
  V.~Andretta\inst{\ref{INAF-OAC}}
  \and
  R.~Susino\inst{\ref{INAF-OATo}}
  \and
  S.~Mancuso\inst{\ref{INAF-OATo}}
  \and
  D.~Spadaro\inst{\ref{INAF-OACt}}
  \and
  M.~Mierla\inst{\ref{ROB},\ref{RomanianAcademy}}
  \and
  D.~Berghmans\inst{\ref{ROB}}
  \and
  E.~D’Huys\inst{\ref{ROB}}
  \and
  A.~N.~Zhukov\inst{\ref{ROB},\ref{SINP}}
  \and
  D.-C.~Talpeanu\inst{\ref{ROB},\ref{KULeuven}}
  \and
  R.~Colaninno\inst{\ref{NRL}}
  \and
  P.~Hess\inst{\ref{NRL}}
  \and
  J.~Koza\inst{\ref{Tatra}}
  \and
  S.~Jej\v{c}i\v{c}\inst{\ref{Lubjana1},\ref{Lubjana2}}
  \and
  P.~Heinzel\inst{\ref{CAS}}
  \and
  E. Antonucci\inst{\ref{INAF}}
  \and
  V. Da Deppo\inst{\ref{CNR-IFN},\ref{INAF}}
  \and
  S. Fineschi\inst{\ref{INAF-OATo}}
  \and
  F. Frassati\inst{\ref{INAF-OATo}}
  \and
  G. Jerse\inst{\ref{INAF-OATs}}
  \and
  F.~Landini\inst{\ref{INAF-OATo}}
  \and
  G. Naletto\inst{\ref{UniPd},\ref{CNR-IFN},\ref{INAF}}
  \and
  G. Nicolini\inst{\ref{INAF-OATo}}
  \and
  M. Pancrazzi\inst{\ref{INAF-OATo}}
  \and
  M.~Romoli\inst{\ref{UniFi},\ref{INAF}}
  \and
  C.~Sasso\inst{\ref{INAF-OAC}}
  \and
  A. Slemer\inst{\ref{CNR-IFN}}
  \and
  M. Stangalini\inst{\ref{ASI}}
  \and
  L. Teriaca\inst{\ref{MPS}}
}

\institute{
  INAF -- Osservatorio Astrofisico di Torino, Turin, Italy\\
  \email{alessandro.bemporad@inaf.it}
  \label{INAF-OATo}
  \and
  INAF -- Osservatorio Astronomico di Capodimonte, 
  Salita Moiariello 16, I-80131 Naples, Italy
  \label{INAF-OAC}
  \and
  INAF -- Osservatorio Astrofisico di Catania, Catania, Italy
  \label{INAF-OACt} 
  \and
  Solar-Terrestrial Centre of Excellence -- SIDC, Royal Observatory of Belgium, Brussels, Belgium
  \label{ROB}
  \and
  Institute of Geodynamics of the Romanian Academy, Bucharest, Romania
  \label{RomanianAcademy}
  \and
  Skobeltsyn Institute of Nuclear Physics, Moscow State University, 119992 Moscow, Russia
  \label{SINP}
  \and
  KU Leuven, Leuven, Belgium
  \label{KULeuven}
  \and
 Naval Research Laboratory, Washington DC, USA
 \label{NRL}
  \and
  Astronomical Institute, Slovak Academy of Sciences, Tatransk\'{a} Lomnica, Slovakia
  \label{Tatra}
  \and
  Faculty of Education, University of Ljubljana, Slovenia
  \label{Lubjana1}
  \and
  Faculty of Mathematics and Physics, University of Ljubljana, Slovenia
  \label{Lubjana2}
  \and
  Astronomical Institute of the Czech Academy of Sciences, Ond\v{r}ejov, Czech Republic
  \label{CAS}
  \and
  CNR -- Istituto di Fotonica e Nanotecnologie, Padua, Italy
  \label{CNR-IFN}  
  \and
  INAF -- Osservatorio Astronomico di Trieste, Trieste, Italy
  \label{INAF-OATs}
  \and
  Universit\`a di Padova -- Dip. Fisica e Astronomia ``Galileo Galilei'', Padua, Italy
  \label{UniPd}  
  \and
  Universit\`a di Firenze -- Dip. Fisica e Astronomia, Florence, Italy
  \label{UniFi}
  \and
  INAF – Associate Scientist
  \label{INAF}
  \and
  Agenzia Spaziale Italiana, Roma, Italy
  \label{ASI}
  \and
  Max-Planck-Institut f\"ur Sonnensystemforschung, G\"ottingen, Germany
  \label{MPS}
}

\date{Received 20/01/2022; accepted 07/02/2022\ldots}

 
\abstract
{On February~12, 2021 two subsequent eruptions occurred above the West limb, as seen along the Sun-Earth line. The first event was a typical slow Coronal Mass Ejection (CME), followed $\sim 7$ hours later by a smaller and collimated prominence eruption, originating Southward with respect to the CME, followed by a plasma blob. These events were observed not only by SOHO and STEREO-A missions, but also by the suite of remote sensing instruments on-board Solar Orbiter (SolO).} 
{This work shows how data acquired by the Full Sun Imager (FSI), Metis coronagraph, and Heliospheric Imager (SoloHI) from the SolO perspective can be combined to study the eruptions and the different source regions. Moreover, we show how Metis data can be analyzed to provide new information about solar eruptions.} 
{Different 3D reconstruction methods were applied to the data acquired by different spacecraft including remote sensing instruments on-board SolO. Images acquired by both Metis channels in the Visible Light (VL) and \ion{H}{i} \Lya\ line (UV) were combined to derive physical information on the expanding plasma. The polarization ratio technique was also applied for the first time to the Metis images acquired in the VL channel.}
{The two eruptions were followed in 3D from their source region to their expansion in the intermediate corona. Thanks to the combination of VL and UV Metis data, the formation of a post-CME Current Sheet (CS) was followed for the first time in the intermediate corona. The plasma temperature gradient across a post-CME blob propagating along the CS was also measured for the first time. Application of the polarization ratio technique to Metis data shows that, thanks to the combination of four different polarization measurements, the errors are reduced by $\sim 5-7$\%, thus better constraining the 3D distribution of plasma.}

  
\titlerunning{A CME followed by prominence eruption and plasma blob observed by Solar Orbiter}
\authorrunning{Bemporad et al.}

\maketitle

\begin{figure*}
  \centering
  \includegraphics[width=0.9\textwidth]{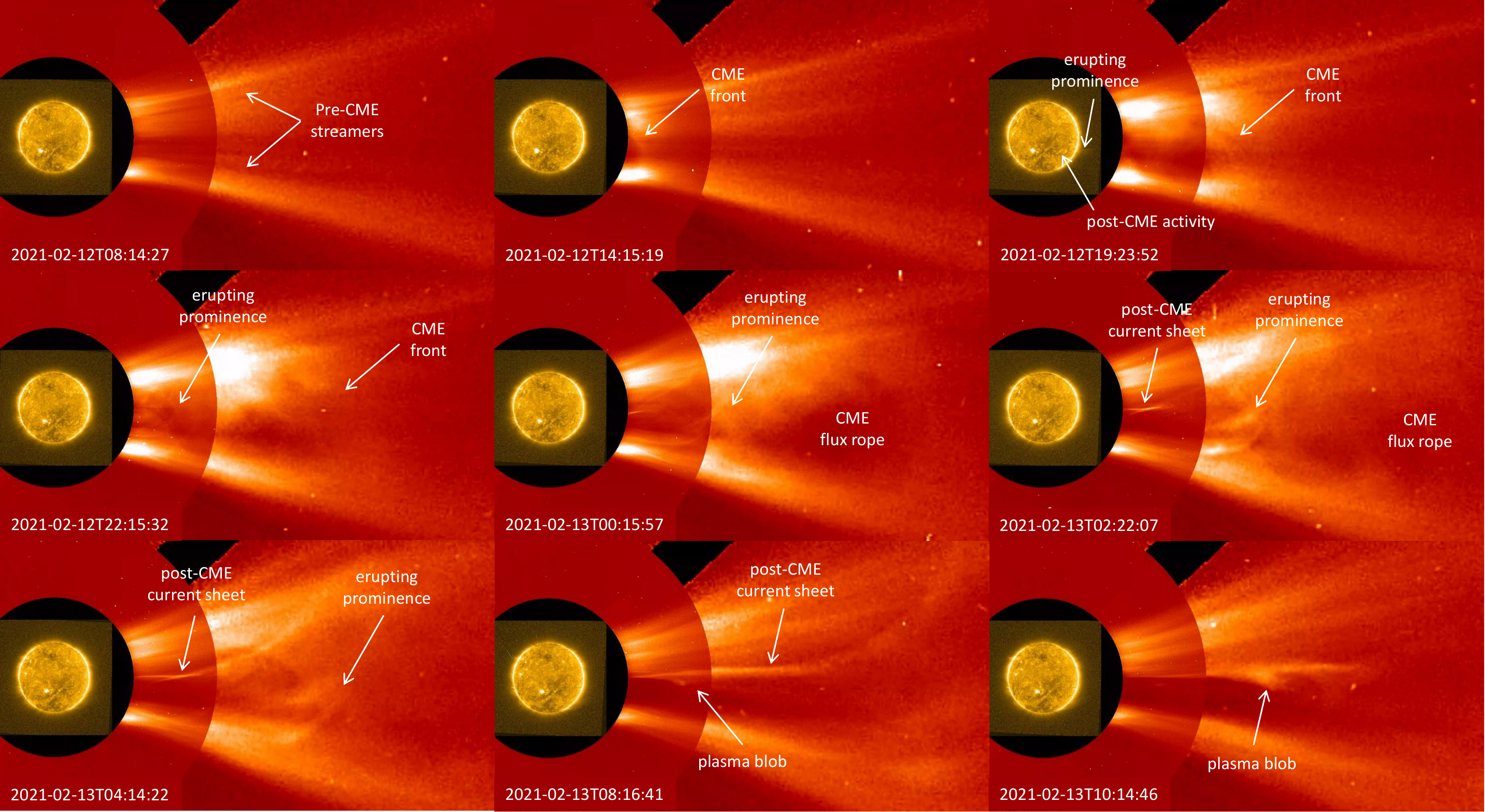}
  \includegraphics[width=0.9\textwidth]{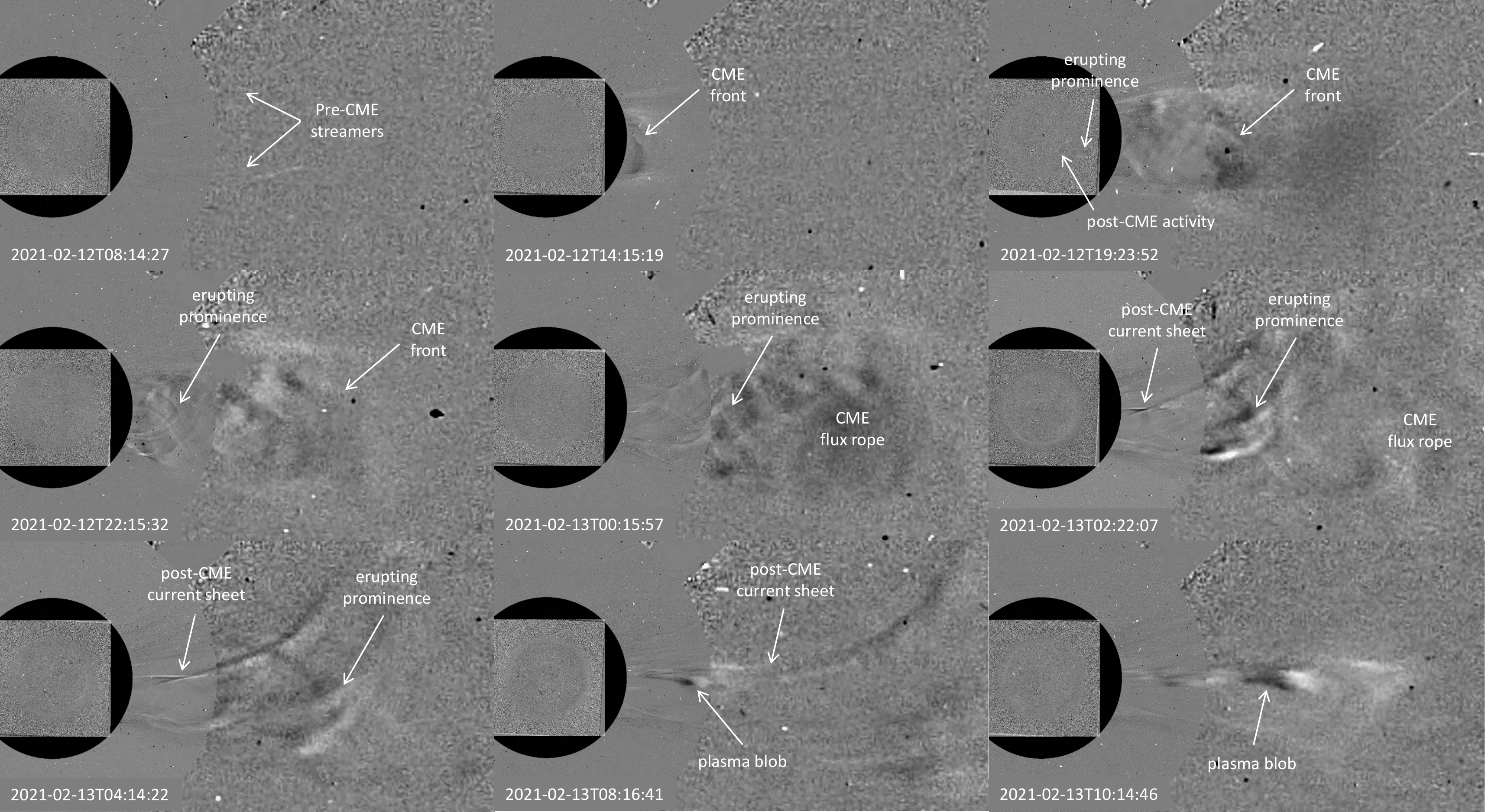}
  \caption{The complex sequence of events that occurred between 2021 February~12 and 13 as shown by the PROBA2/SWAP imager and SOHO/LASCO-C2 and C3 coronagraphs (see text). The top part of the Figure shows intensity images, while the bottom part of the Figure shows the corresponding running difference images. All these combined images have been created with JHelioviewer \citep{Mueller-etal:2017}.}
  \label{Fig:CMEdescription}
\end{figure*}

\section{Introduction}

Space-based coronagraphs are the main tools available so far to continuously monitor the intermediate solar corona \citep[see e.g.][]{Hochedez-2005}. Images acquired by these instruments provide unique input for the release of the first alert and forecasting on the occurrence of prominence eruptions and Coronal Mass Ejections (CMEs) and their possible impact on the Earth’s magnetosphere \citep[see review by][]{Webb-Howard:2012}. Over the last decades, many space-based coronagraphs allowed us to continuously monitor CMEs and to study their early evolution from $\sim 1.5$ to $\sim 30$\,\rsun\ over more than two solar cycles \citep[][]{Yashiro-etal:2004,Gopalswamy-etal:2009}, and more than $4 \times 10^4$ CMEs have been observed by these instruments \citep{Lamy-etal:2019}. The more recent multi-spacecraft coronagraphic observations provided by the STEREO mission also allowed (in combination with data acquired by other instruments) to investigate and unveil the 3D structure of CMEs, which can be now reconstructed with many different forward and backward modeling techniques \citep[see review by][and references therein]{Thernisien-2011}.

With the launch of the Solar Orbiter spacecraft \citep[SolO,][]{Mueller-etal:2020} in February 2020, a new era of the space-based coronagraphy started: the era of multi-band coronagraphy. In particular, the Metis coronagraph \citep{Antonucci-etal:2020} on-board SolO is the first instrument observing at the same time in the polarised Visible Light (VL) broad-band in the interval 580-640\,nm, and UV narrow-band centered around the \SI{121.6}{\nano\meter} \Lya\ line emitted by neutral H atoms (the most intense line in the UV solar spectrum). Combined analysis of VL and UV data can provide 2D maps of the electron density and temperature inside CMEs \citep{Bemporad-etal:2018}, thus allowing to study for the very first time the thermo-dynamical evolution of CMEs during their early expansion phases \citep[e.g.][]{Susino-Bemporad:2016}. These kinds of observations could be potentially decisive to finally understand the physical origin of the mysterious additional heating source observed by the UVCS spectrometer \citep[UV Coronagraph Spectrometer;][]{Kohl-etal:1995} on-board SOHO and reported by many previous authors \citep[see][and references therein]{Wilson-2021}. A comprehensive review of CME observations with UVCS can be found in \cite{Kohl-etal:2006}, while a catalogue is given by \cite{Giordano-etal:2013}.

Unlike coronagraphs, space-based EUV-XUV imagers are the only instruments capable of providing unique information about the pre-CME evolution in their source regions and their early evolution in the inner corona \citep{Georgoulis-2019}, as well as the evolution of related phenomena \citep{Hudson2001,Zhukov2007}, such as the so-called EUV waves \citep[see review by][]{Liu2014}, EUV dimmings \citep[e.g.][]{Zhukov-2004}, post-eruption arcades \citep{Tripathi2004} and CME-driven shocks \citep[see e.g.][]{Shen-2012,Mancuso2019}, thus helping in the identification of the most probable source of type II radio bursts and acceleration of Solar Energetic Particles (SEPs) \citep[e.g.][]{Frassati-2019, Kozarev-2015}. These instruments were used for a long time, in combination with photospheric magnetometers, to identify possible precursors or progenitors of CMEs and flares \citep[see review by][]{Chen-2011}.
Moreover, since the launch of the STEREO mission,
the CME origins in the inner EUV corona can now also be imaged from different perspectives than the Sun-Earth line. The Full Sun Imager (FSI), a telescope of the EUI instrument \citep{Rochus-2020} on-board SolO, added an extra viewpoint with images in its \SI{17.4}{\nano\meter} and \SI{30.4}{\nano\meter} channel.  A particularly useful aspect of FSI for CME studies is its large 3.8\degr\ field of view (FoV), which provides ample overlap with the Metis FoV \citep[see][]{2020A&A...642A...6A}.

To track CMEs further away from the Sun, space-based Heliospheric Imagers such as SMEI \citep{Jackson-2004} and STEREO/HI \citep{Howard-2008} became a fundamental platform to understand the Sun-to-Earth propagation of any solar disturbance, such as the Interplanetary counterpart of CMEs (ICMEs). These instruments became the baseline for driving our understanding of the evolution of ICMEs and their associated shocks, as well as for the prediction of their arrival times on Earth and their possible geo-effectiveness \citep[see recent review by][]{Temmer-2021}.

This capability is now being further enhanced thanks to the SolO Heliospheric Imager \citep[SoloHI;][]{Howard-2020}. SoloHI is comprised of four separate Advanced Pixel Sensor (APS) detectors combined to form a single optical system observing the heliosphere in white light. The instrument has a 40\degr\ FoV, typically centered near 0\degr\ in latitude and 25\degr\ elongation from the Sun. SoloHI is on the anti-ram side of the spacecraft, meaning that in a nominal observing configuration for the spacecraft SoloHI is always observing off the East limb of the Sun.

\begin{figure}
  \centering
  \includegraphics[width=0.49\textwidth]{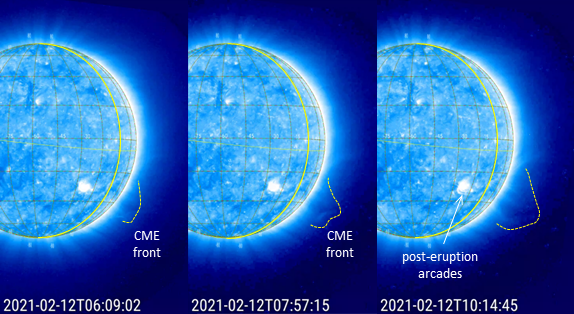}
  \caption{The early expansion of the CME front as observed at the beginning of February~12 in the STEREO-A/EUVI \SI{17.1}{\nano\meter} channel (see text). The yellow solid line shows the location of the central meridian as seen from Earth, while the latitude and longitude grid coordinates are given in the Stonyhurst reference frame.}
  \label{Fig:EUVIfront}
\end{figure}

In this paper, the observations of three subsequent solar eruptions acquired by the above remote sensing instruments on-board SolO are analyzed. After a description of the events as observed by well established instruments to introduce the main on-going phenomena (Sec.~\ref{sec:general}), we describe in more details the images acquired by the EUI (Sec.~\ref{sec:eui_obs}), Metis (Sec.~\ref{sec:metis_obs}), and SoloHI (Sec.~\ref{sec:hi_obs}) instruments. The combination of these data allowed us to determine the 3D kinematics of the eruptions, as discussed in Sec.~\ref{sec:3d_kinematics}. Finally (Sec.~\ref{sec:metis_polariz}), the Metis polarized images acquired with the VL channel are analyzed and inverted with the so-called polarization ratio technique \citep{Moran-2004}. Results are then summarized in the conclusions (Sec.~\ref{sec:conclusions}) also discussing the advantages of the Metis coronagraph with respect to previous similar instruments.
\begin{figure}
  \centering
  \includegraphics[width=0.49\textwidth]{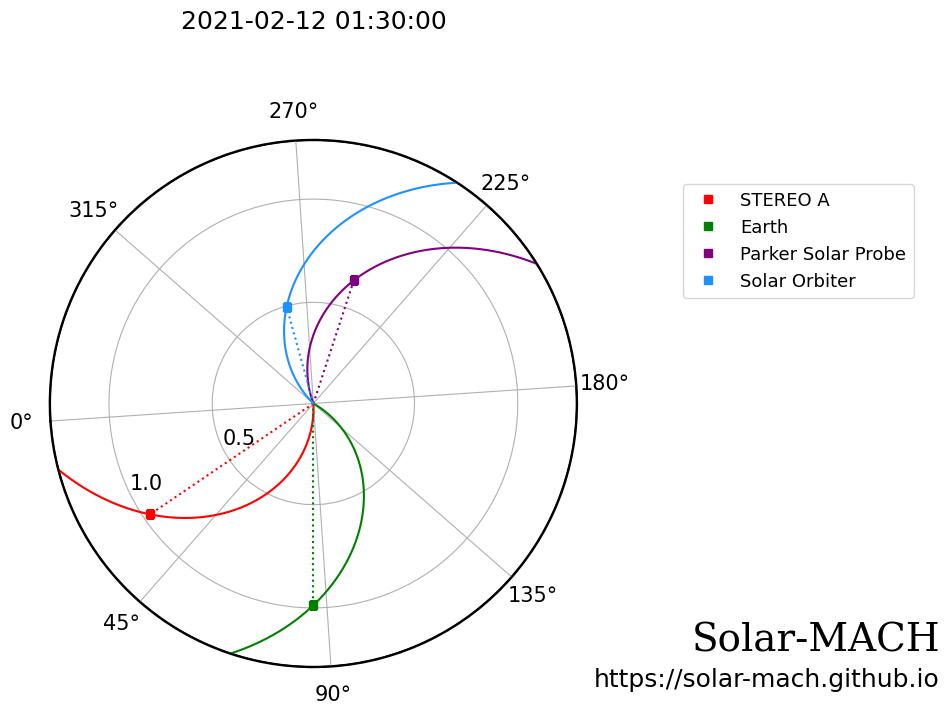}
  \caption{Locations of the different spacecraft and the Earth between February~12--13, 2021 as seen projected onto the ecliptic plane.}
  \label{Fig:spacecraft}
\end{figure}

\section{Observations from the Sun-Earth line and STEREO-A}
\label{sec:general}

On February~12, 2021 two subsequent eruptions occurred above the West limb, as seen along the Sun-Earth line. The complex sequence of events is shown in Fig.~\ref{Fig:CMEdescription}, both with regular (top) and running difference (bottom) images. The first event appeared in the SOHO/LASCO-C2 images \citep{Brueckner-etal:1995} as a typical CME, starting to appear in the instrument FoV around 12:48\,UT. The CME propagated in between two pre-existing coronal streamers, mostly unaffected by the eruption. The subsequent images (see Fig.~\ref{Fig:CMEdescription}) show the progressive expansion of the CME front enclosing a flux-rope like structure without any clear CME core. The CME had a projected speed on the order of $\sim 110$\,\kms, as provided by the CACTUS catalog \citep{Robbrecht-Berghmans:2004}. More refined evaluations, based on LASCO running difference images, give values on the order of $\sim 80-90$\,\kms\ in the LASCO-C2 FoV and $\sim 110$\,\kms\ when the CME front is entering in the LASCO-C3 FoV, hence this can be classified as a slow and accelerating event, as also reported by the CDAW CME catalog \citep{Gopalswamy-etal:2009} providing an average acceleration by 5.8\,m\,s$^{-2}$.

\begin{figure*}
  \includegraphics[width=\textwidth]{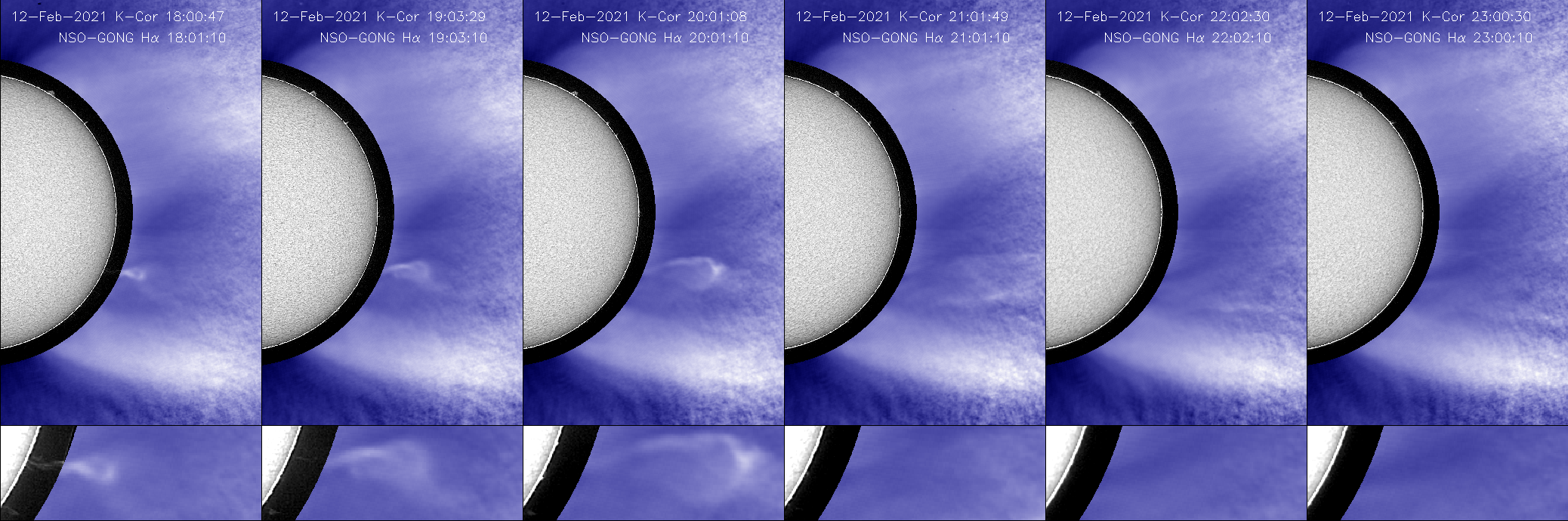}
  \caption{The prominence eruption as observed in the late hours of February~12 by the K-Cor coronagraph and the NSO-GONG network H$\alpha$ monitor at the Mauna Loa Solar Observatory (see also Fig.~\ref{Fig:FSIeruption}). The size of FoV of the top and bottom panels are $1840\arcsec \times 3000\arcsec$ and $850\arcsec \times 310\arcsec$, respectively. The outer radius of the off-limb annulus is about 115\arcsec\ larger than the solar radius.}
  \label{Fig:MLSOprominence}
\end{figure*}

\begin{figure}
  \includegraphics[width=0.49\textwidth]{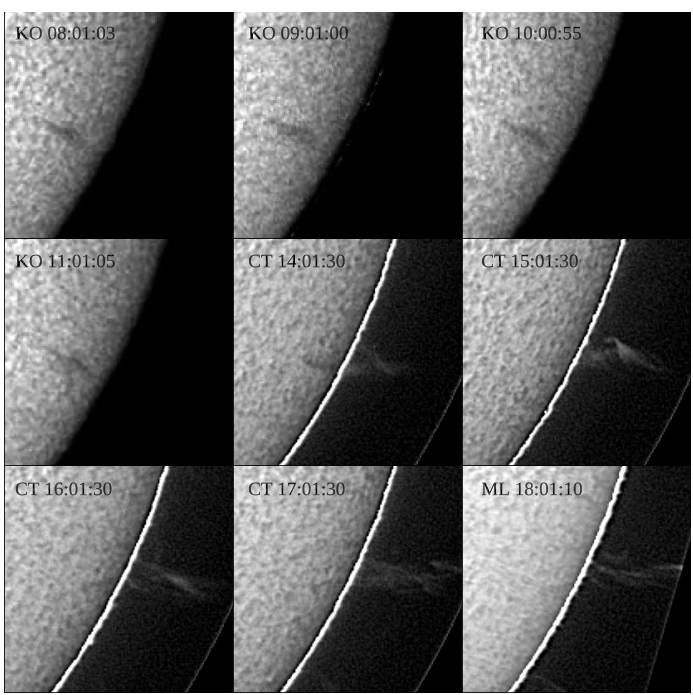}
  \caption{The prominence eruption as observed from the beginning by the Kanzelh\"{o}he Observatory (KO) and the NSO-GONG network H$\alpha$ monitors at the Cerro Tololo (CT) and the Mauna Loa Solar Observatory (ML). The size of FoV is $300\arcsec \times 300\arcsec$.}
  \label{Fig:GONGprominence}
\end{figure}
Accordingly, the CME was not associated with any solar flare, filament eruption, or radio burst, but the EUV images acquired by the PROBA2/SWAP instrument \citep{Berghmans-2006, Seaton2013} show some activity (formation of post-eruption arcades) going on in a region located around 30\degr-40\degr\ West and 30\degr\ South (in the Stonyhurst coordinate system) starting from $\sim 10$\,UT, hence two hours before the appearance of the CME in the LASCO FoV, and outlined in Fig.~\ref{Fig:EUVIfront} (third panel). This is the most likely location of the CME source region, as also supported by the EUV full disk images acquired by the STEREO-A/EUVI instrument \citep{Howard-2008} in the \SI{17.1}{\nano\meter} channel. In particular, on February~13, 2021 at 00:00\,UT the STEREO-A spacecraft was located at a separation angle with the Earth of 55.86\degr\ East (see Fig.~\ref{Fig:spacecraft}), and from this vantage point the spacecraft observed the off-limb early propagation phase of the CME, appearing at 10:09\,UT as a loop arcade expanding from the West limb close to the instrument plane-of-sky with the two footpoints located around $\sim 20\degr$ South and $\sim 40\degr$ South (see Fig.~\ref{Fig:EUVIfront}, right panel) in the Stonyhurst coordinates.

This slow CME was followed by a complex sequence of two small-scale events. First, also observed from the ground by the K-Cor coronagraph at the Mauna Loa Solar Observatory and the Global Oscillation Network Group H$\alpha$ network monitor operated by the National Solar Observatory, \citep[NSO-GONG,][]{Harveyetal2011}, a prominence following the CME started to erupt on February~12 around 18:00\,UT, leaving the instrument FoV around 21:00\,UT (see Fig.~\ref{Fig:MLSOprominence} and relevant data\footnote{\url{https://search.datacite.org/works/10.5065/d69g5jv8}}). The prominence appeared to leave the solar limb as seen from the Earth from a projected latitude of about 25\degr~S, propagating mostly toward the equatorial plane. The initial projected propagation speed of the prominence is low, on the order of 22 \kms\ as measured with K-Cor images, but the prominence is slowly accelerating up to a speed of about 51 \kms\ when it is leaving the instrument FoV. During its early expansion phase, the prominence clearly assumed a hook-like shape, as shown in detail by the third bottom panel in Fig.~\ref{Fig:MLSOprominence} acquired at 20:01\,UT. This apparent shape is likely related with twisting motions occurring in the early expansion phases: the H$\alpha$ images, acquired at earlier times at the Kanzelh\"{o}he Observatory \citep{Potzietal2015} and by the NSO-GONG network monitors (Fig.~\ref{Fig:GONGprominence}), show that the prominence rotated around a vertical axis during the outward propagation, changing its orientation from mostly East-West to mostly North-South. Interestingly, the K-Cor images also show that starting from $\sim 21$\,UT the northern leg of the prominence leaves the Sun resulting in an almost radial intensity enhancement, while its southern leg material is apparently deposited in the surrounding corona.

Associated with the prominence, the PROBA2/SWAP instrument also detected at the same time an expanding tongue of plasma, outlined in Fig.~\ref{Fig:CMEdescription} (top right panel), also observed by the SDO/AIA telescopes. Hence, higher up in the LASCO-C2 FoV, this eruption resulted in more filamentary plasma features following the CME (Fig.~\ref{Fig:CMEdescription}, middle row panels). The clear identification of features observed in the coronagraphs with the prominence is complicated by the occurrence of the nearby CME partially aligned along the line-of-sight (LoS), and also because of the described prominence twisting motions observed in the low corona and probably proceeding higher up during the expansion. The prominence, propagating at about 80\,\kms in the LASCO FoV, finally merged with the post-CME plasma, becoming nearly indistinguishable from the CME in the LASCO-C3 FoV. The observations of the same prominence by SolO/FSI will be discussed below.

\begin{figure*}
  \centering
  \includegraphics[width=0.325\textwidth]{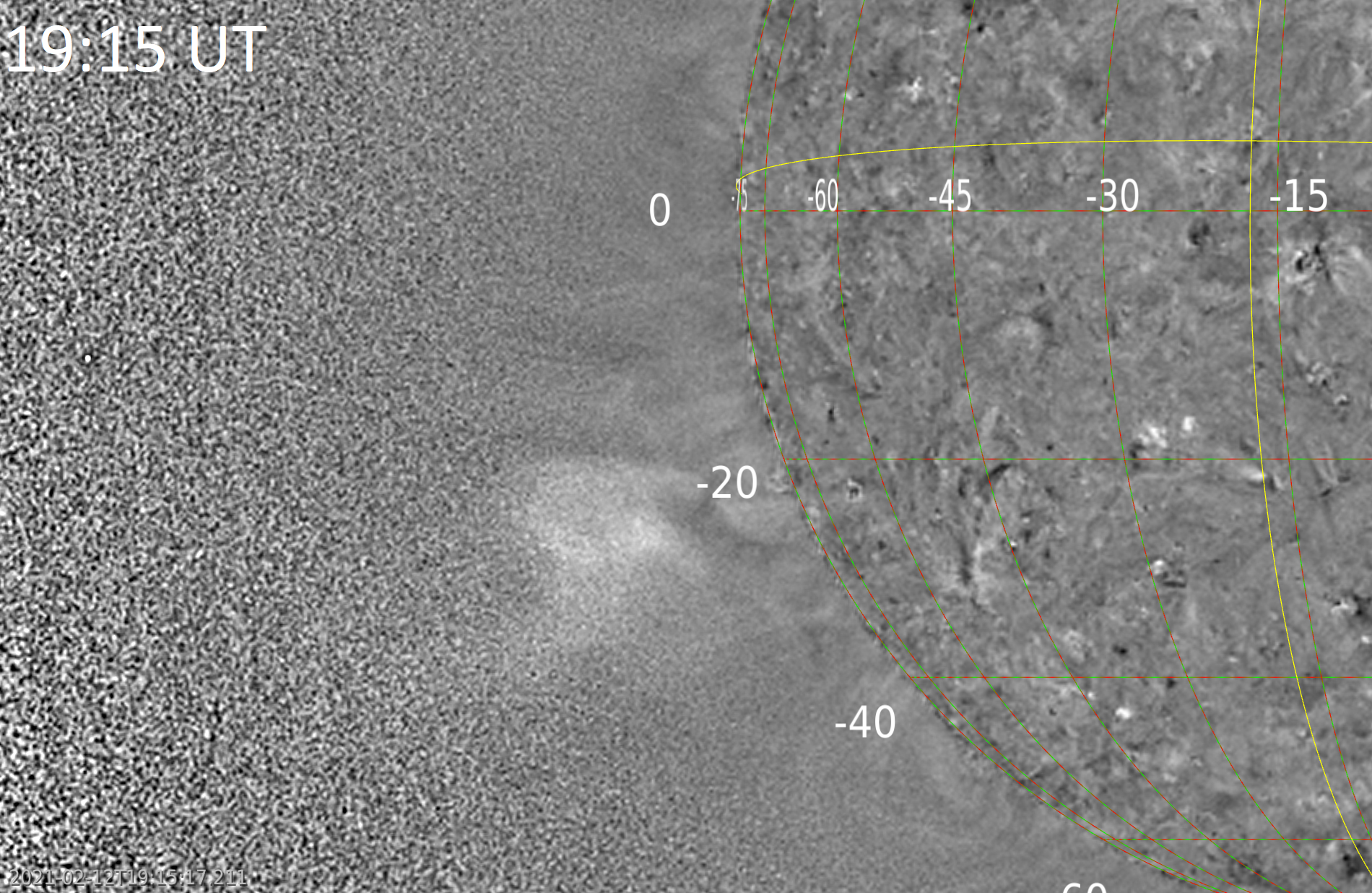}
  \includegraphics[width=0.325\textwidth]{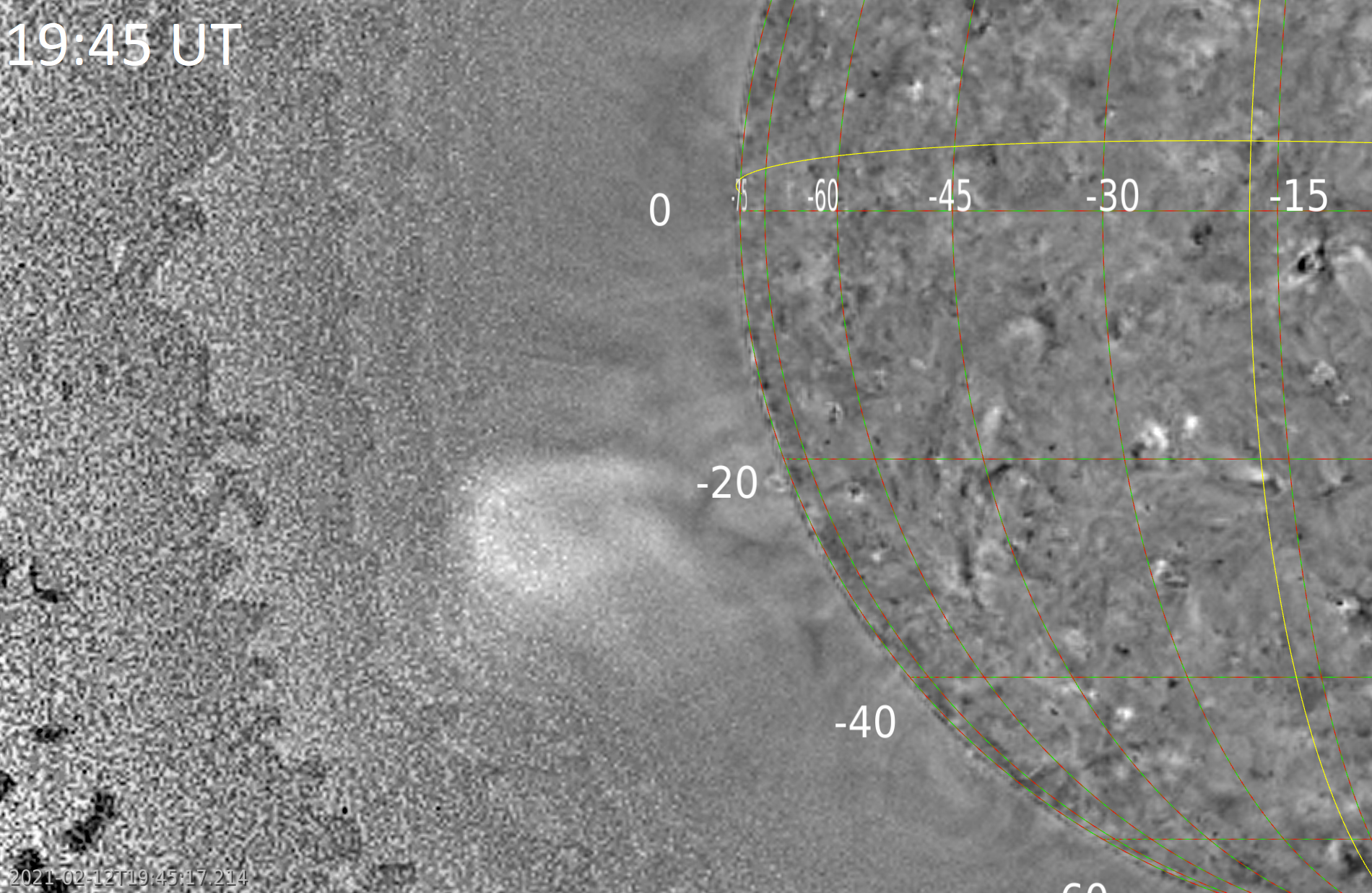}
  \includegraphics[width=0.325\textwidth]{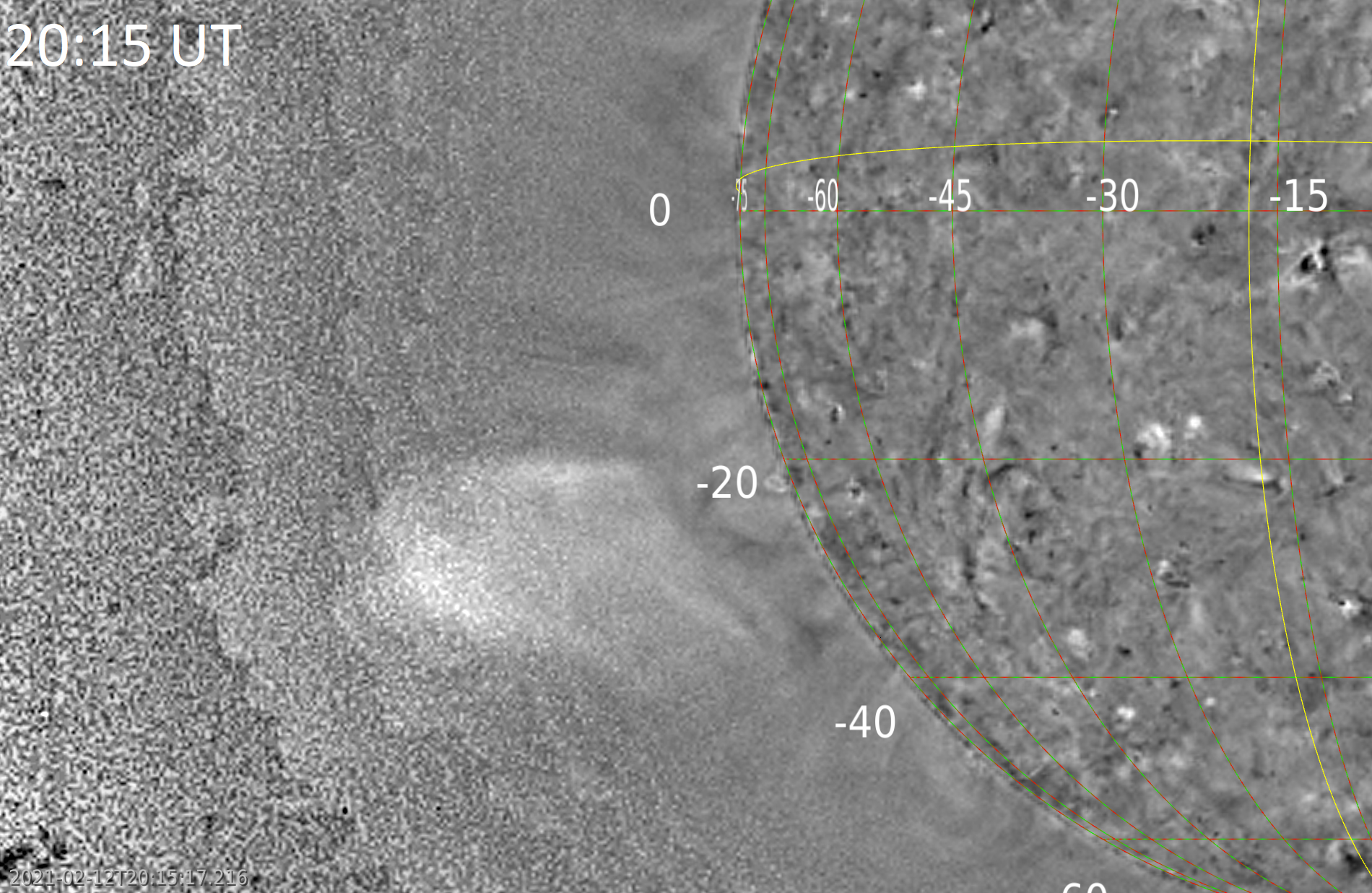}
  \includegraphics[width=0.325\textwidth]{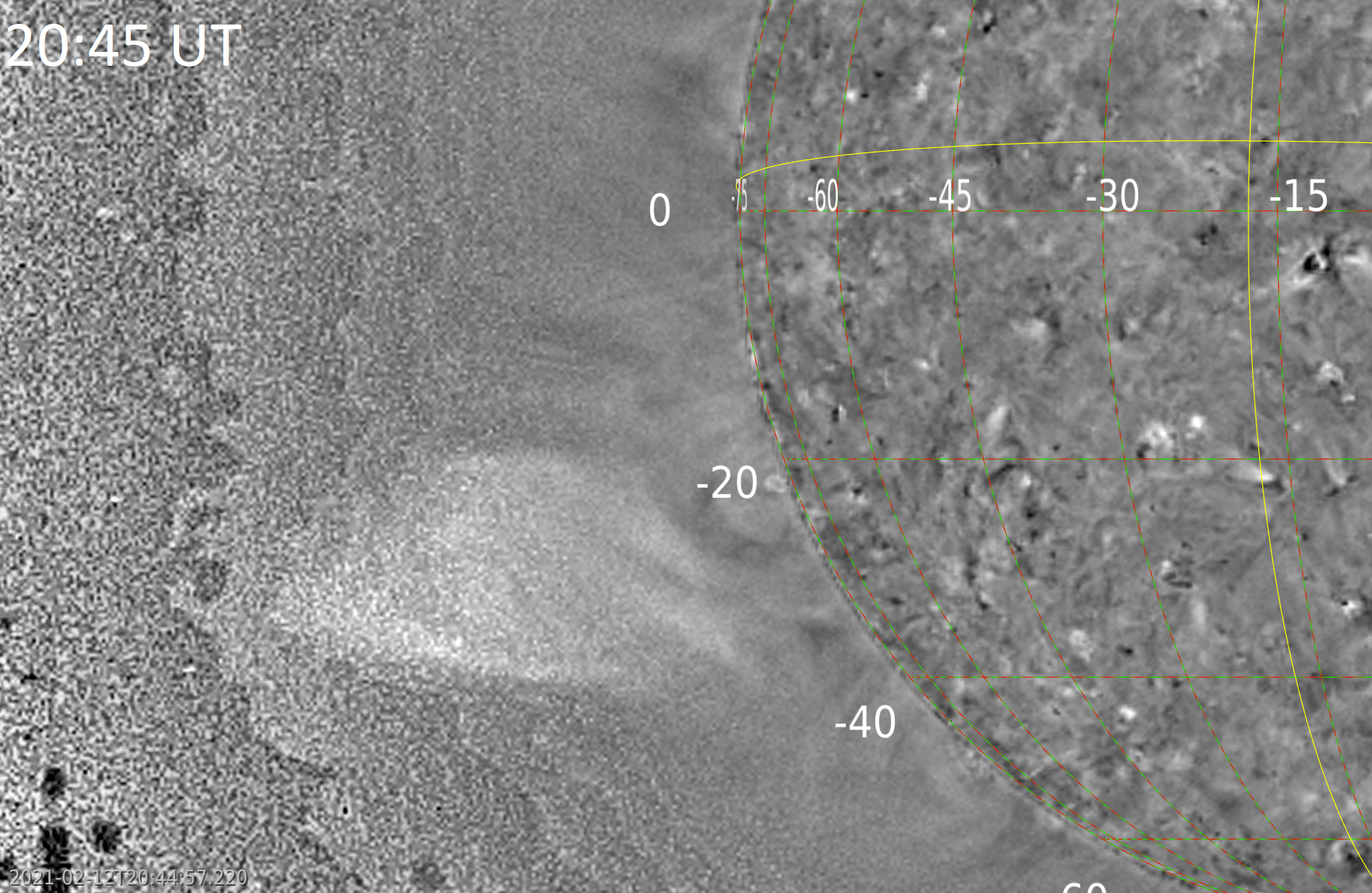}
  \includegraphics[width=0.325\textwidth]{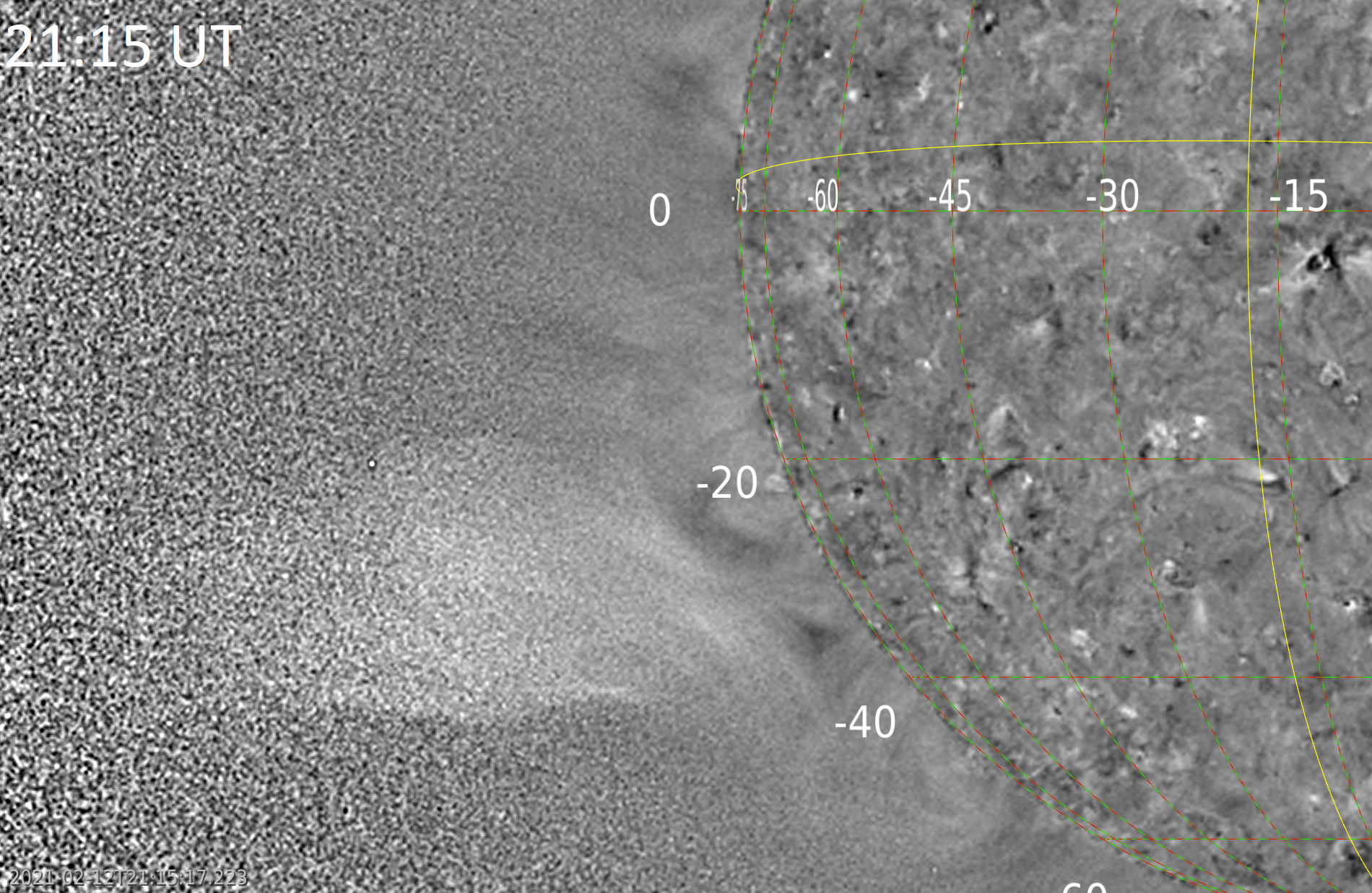}
  \includegraphics[width=0.325\textwidth]{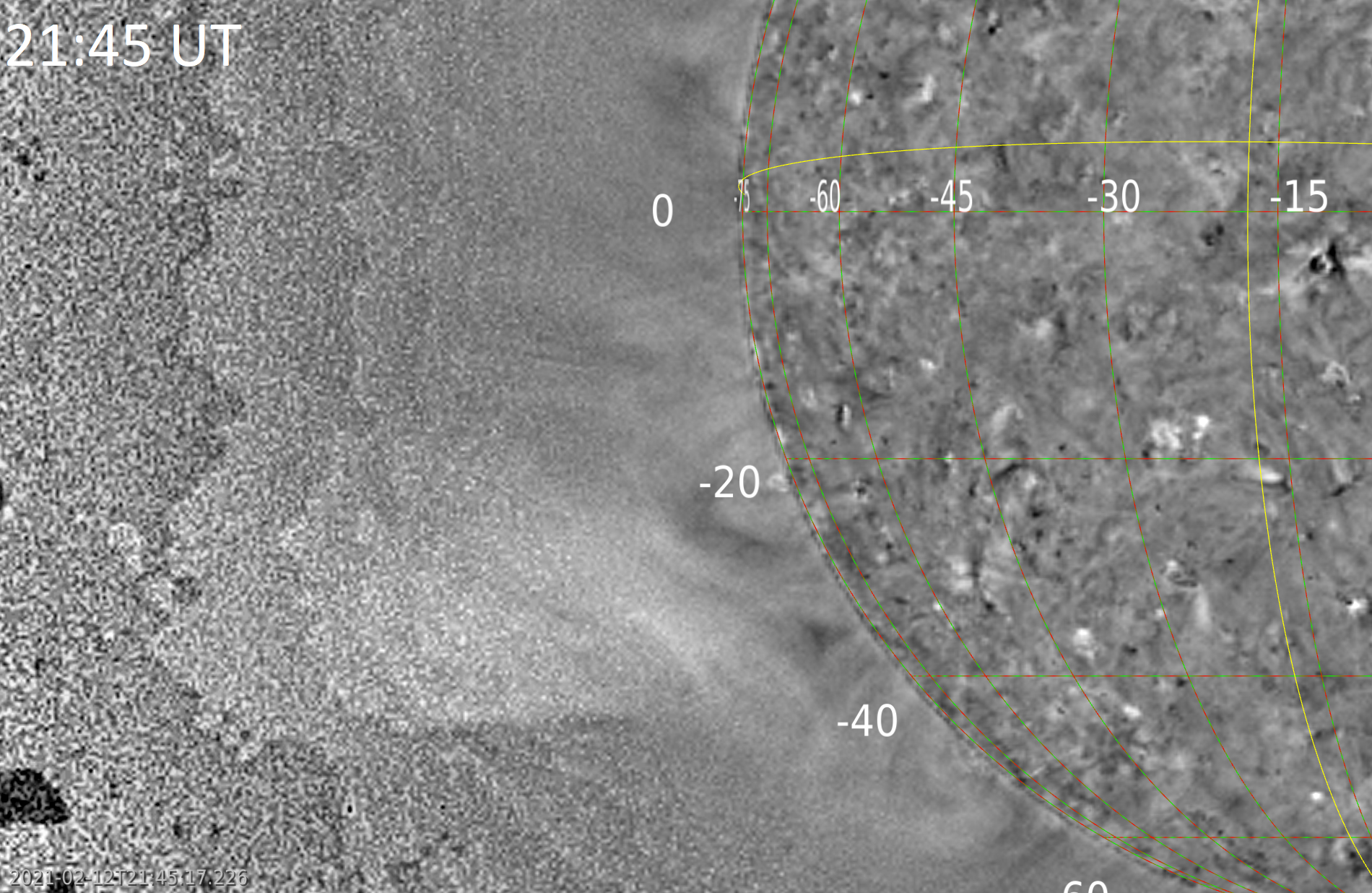}
  \caption{Eruption of a bright structure (corresponding to the prominence shown in Fig.~\ref{Fig:MLSOprominence}) observed in FSI \SI{17.4}{\nano\meter} base difference images taken on 2021 February~12 at the different times given in each panel. The image taken at 15:15\,UT was subtracted from each of the original images to reveal the evolution of the coronal structure during the eruption. An online movie showing the event is also available. The images and the movie were created with the JHelioviewer software \citep{Mueller-etal:2017}.}
  \label{Fig:FSIeruption}
\end{figure*}
Then, after the CME and the prominence, the LASCO images show the propagation of an ``inverted V-shape'' smaller-scale intensity enhancement (Fig.~\ref{Fig:CMEdescription}, middle row panels), resulting in a radial coronal feature (aligned with the CME propagation latitude) that we identify here as the post-CME Current Sheet (CS). This identification will be supported and discussed below thanks to the Metis observations. Moreover, after the CME and the prominence, the LASCO images also show the expansion of a collimated plasma blob, almost co-aligned with the post-CME CS, and propagating much faster at a projected speed of $\sim 380$\,\kms\ (see Fig.~\ref{Fig:CMEdescription}, middle and right bottom panels) in the LASCO/C2 FoV, thus rapidly merging later on with the slow CME and prominence propagating ahead. The origin of this post-CME blob is more uncertain because no clear activity was observed by the SDO/AIA and PROBA2/SWAP full-disk imagers. The study of this blob with SolO/Metis data will also be discussed below.

Considering the small separation angle between the apparent source regions of the CME (Fig.~\ref{Fig:CMEdescription}, top panels) and the prominence eruption (Fig.~\ref{Fig:MLSOprominence}) and their time sequence, these two events could be candidates for sympathetic eruptions \citep[see review by][]{Lugaz-2017}, with the first one (a typical slow CME with a front and a cavity, but no clear core) destabilizing a nearby prominence and ``opening'' (by some kind of interchange reconnection) the route for its propagation, leading finally to the prominence eruption expanding near the flanks of the CME. A similar sequence of events was described for instance by \citet{Bemporad-2008}. The subsequent formation of a post-CME CS (Fig.~\ref{Fig:CMEdescription}, middle panels) is typical for solar eruptions, because these features are believed to form as a consequence of post-CME magnetic reconnection \cite[e.g.][]{Lin_2000}. This also suggests a possible physical explanation for the subsequent more compact and faster plasma blob (Fig.~\ref{Fig:CMEdescription}, bottom panels). Because this feature was also apparently aligned with the prominence, it is possible that the blob is accelerated by magnetic reconnection going on in the reconfiguration phases of solar corona in the post-CME CS. The formation of similar plasma blobs is typical for the evolution of post-CME CS \cite[e.g.][]{Ko_2003, Vrsnak_2009}, and is believed to be driven by tearing instability related with the CME expansion \cite[e.g.][]{Shibata2001, Barta_2008}.

\section{The events as observed from Solar Orbiter}
\label{sec:solo}

In order to understand the different appearance of the events described above as seen from SolO, it is important to consider that between February~12--13, 2021 the spacecraft was separated by 161.6\degr\ East from the Earth in heliographic longitude (see Fig.~\ref{Fig:spacecraft}), hence was mostly observing the opposite solar hemisphere with respect to the Earth. This means that all the events that were observed above the West (East) limb from the Sun-Earth line view, are expected to propagate above the East (West) limb as seen from SolO.

\subsection{EUI observations}
\label{sec:eui_obs}

In this paper, we use the Release 2 of the calibrated EUI/FSI data\footnote{\url{https://doi.org/10.24414/z2hf-b008}}.
The cadence of the FSI \SI{17.4}{\nano\meter} images was around 30~minutes (see Fig.~\ref{Fig:FSIeruption}), and no \SI{30.4}{\nano\meter} images were taken. There is no counterpart of the main slow CME observed in the Metis data, probably due to the position of its source region on the opposite side of the Sun (see Section~\ref{sec:general}), and the low speed of the CME, which does not allow seeing the development of coronal dimmings above the limb \citep{Kilpua2014,Palmerio2021}. The eruption of the prominence was, on the contrary, well observed by FSI (Fig.~\ref{Fig:FSIeruption}). The corresponding structure (bright in the \SI{17.4}{\nano\meter} passband) above the East limb is slowly rising starting from around 16:15\,UT on February~12. By 18:15\,UT a concave-out structure is formed, which evolves to the hook-like morphology seen at 19:45\,UT. The hook-like structure rises at the average speed of $\sim 40$\,\kms\ and around 22:15\,UT it becomes difficult to distinguish among compression artefacts in the outer FoV of FSI. The southern leg of the structure can still be seen slowly rising even early on February~13, while the northern leg is already erupted. No clear coronal dimming was observed. 
\begin{figure*}
  \centering
  \includegraphics[width=0.75\textwidth]{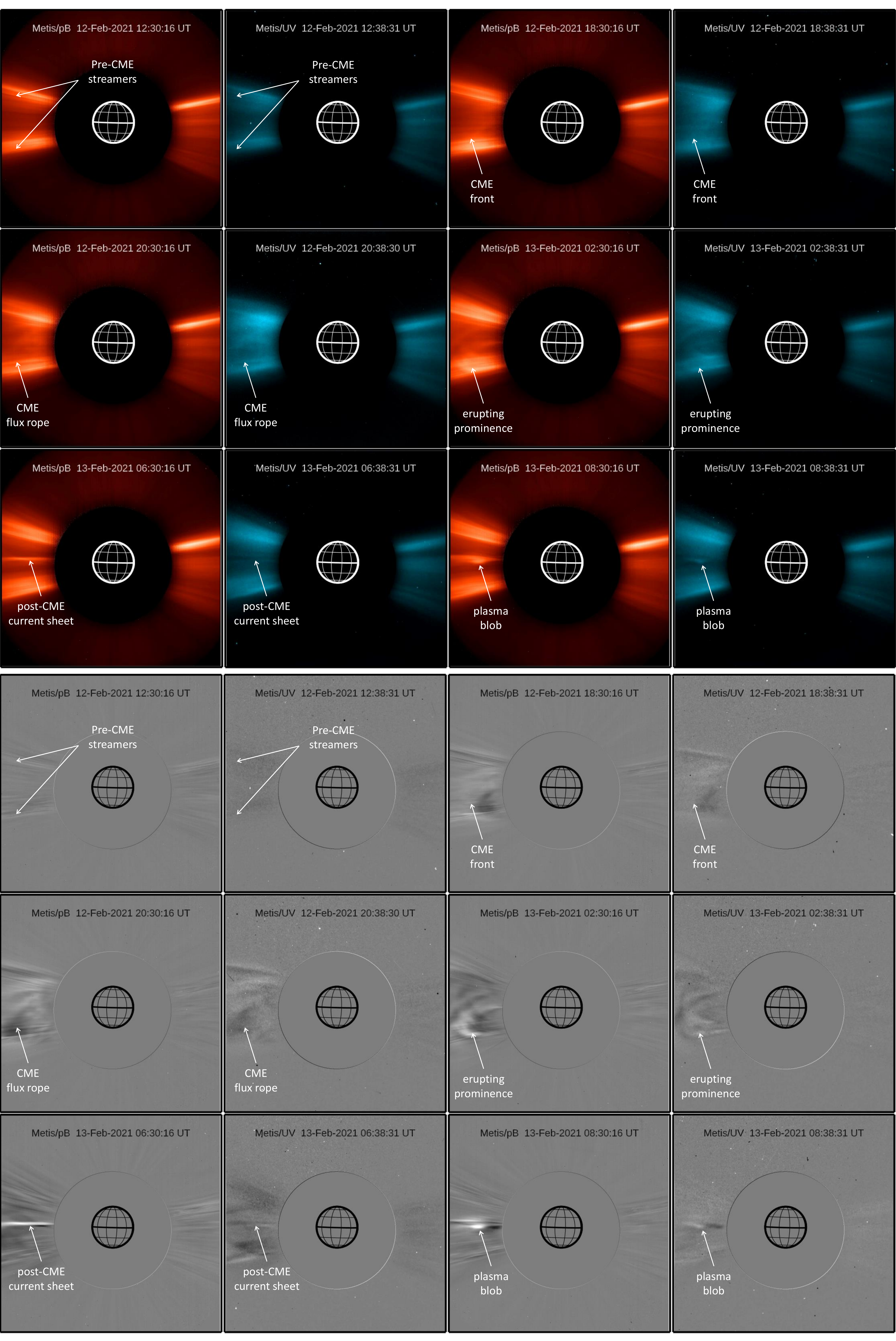}
  \caption{The sequence of events that occurred between February~12--13, 2021 as observed by the Metis coronagraph in the VL (red) and UV (cyan) channels (see text). The top part of the Figure shows regular intensity images, while the bottom part of the Figure shows the corresponding running difference images.}
  \label{Fig:METISimages}
\end{figure*}

The origin of the propagating plasma blob observed by LASCO at 08:16\,UT (Fig.~\ref{Fig:CMEdescription}, bottom middle panel) is not seen in the low corona by FSI. As anticipated, this structure was probably formed higher up in the corona in the post-CME CS. 
\begin{figure*}
  \centering
  \includegraphics[width=\textwidth, bb = 5 5 500 240]{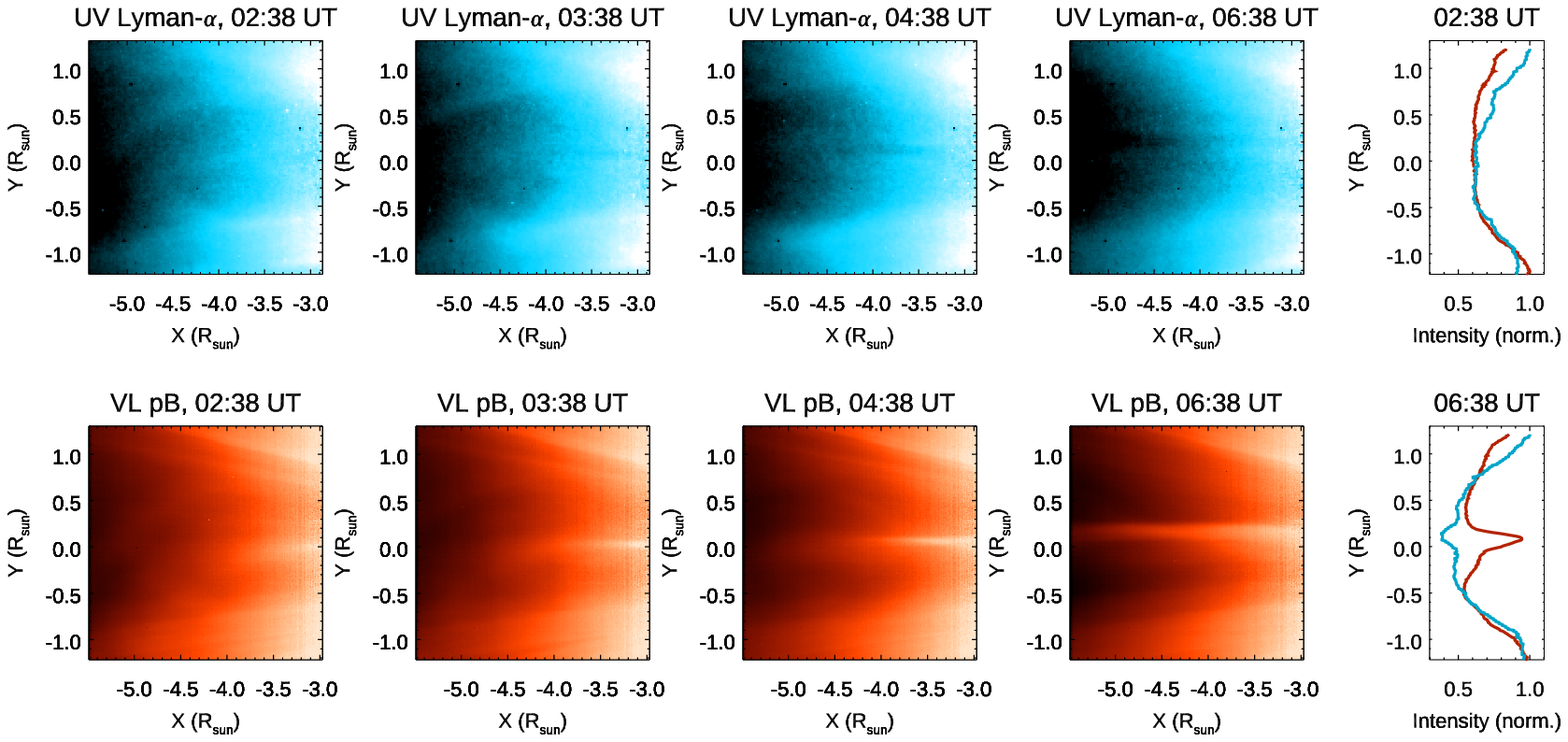}
  \caption{\textit{Left:} a zoom over the post-CME Current Sheet (CS) as observed by Metis on February~13 between 02:38 and 06:38\,UT in the UV (top) and the VL (pB, bottom) channels (see also bottom right panel of Fig.~\ref{Fig:METISimages}). \textit{Right:} normalized UV (blue) and VL (red) North-South intensity profiles as obtained at 02:38 UT (top) and 06:38 UT (bottom) by averaging East-West between $-5.4$ \rsun\ and $-4.4$ \rsun\ (see text for explanations).}
  \label{Fig:MetisCS}
\end{figure*}

\subsection{Metis observations}
\label{sec:metis_obs}

Between 2021 February~12--13, the Metis coronagraph acquired low-cadence sequences of VL and UV synoptic images. Both instrument detectors were binned with a $2\times2$ binning, resulting in $1024 \times 1024$\,pix$^2$ VL images and $512 \times 512$\,pix$^2$ UV images, with spatial scales of 20.3\arcsec\,pix$^{-1}$ and 40.8\arcsec\,pix$^{-1}$, respectively. Considering that the SolO spacecraft was located at a heliocentric distance of $\sim 0.5$\,AU, the Sun was observed with an apparent photospheric radius of 1934.52\arcsec\ and the above projected plate scales correspond on the instrument plane of the sky to about 7300\,km\,pix$^{-1}$ in the VL and 15000\,km\,pix$^{-1}$ in the UV.

The instrument acquired one full VL polarimetric sequence every hour, consisting of four images at different orientations of the linear polarizer, in parallel with two UV images. Each VL image was obtained by averaging on-board 15 frames acquired at the same polarizer orientation with integration time of 30\,s, corresponding to a total effective exposure time of 450\,s. However, since Metis polarimetric observations are carried out cycling over the four specified polarizer orientations for each frame acquired, resulting in an interleaved stream of polarimetric frames \citep[see Sect. 9.1 in][]{Antonucci-etal:2020}, the time elapsed between the acquisition of the first and the last frames corresponding to the same polarizer orientation is of about 30~minutes, i.e. nearly equal to the time taken by the acquisition of the full sequence. The UV images were obtained by averaging 15 frames acquired with integration time of 59.967\,s, corresponding to a total effective exposure time of 899.5\,s, so that the acquisition of both images required about 30~minutes.

The data were calibrated by performing standard operations such as bias and dark subtraction, flat-field and vignetting corrections, and exposure-time normalization \citep[see][]{Romoli-etal:2021}. The improvements in the UV vignetting function described in Appendix~A of \citet{Andretta-2021} were also included. Each set of polarimetric VL images were combined together using the M\"{u}ller matrix obtained from laboratory calibrations to derive the corresponding total- (tB) and polarized-brightness (pB) images.

It is worth noting that the synoptic program to which the observations presented here belong was primarily designed to monitor the status of the solar corona in a long time interval and to provide additional context data for the joint science with the other SolO instrumentation, thus acquisitions were not optimized for the study of solar transients and CMEs. For this reason, considering that the acquisition of each VL sequence was approximately twice as long as that of any single UV image, some differences are expected in the appearance of moving features. In particular, given the above pixel scales and acquisition times, a partial blurring in the images is expected for every feature moving faster than $\sim 4$\,\kms\ and $\sim 16$\,\kms\ in the VL and UV channels, respectively. In order to take into account the different duration of the UV and VL acquisitions, it is possible to average the two subsequent UV images.

Selected VL pB and UV images are shown in the different panels of Fig.~\ref{Fig:METISimages} (top), in red and cyan colors respectively; for each image, the starting acquisition time is given at the top. Similar to Fig.~\ref{Fig:CMEdescription}, also this Figure provides the identification of different features in the regular (top) and running difference (bottom) images. This Figure shows that Metis observed the first CME as a very faint front enclosing a darker cavity and followed by multiple filamentary features without any evident core. The CME appeared to propagate in the Metis FoV above the East solar limb as seen from SolO, moving at an apparent projected speed on the order of $\sim 80 - 100$\,\kms\ (a more refined value of this speed will be provided later based on 3D reconstructions). After the transit of the CME, the Metis images also show faint filamentary expanding features that resembles those also observed by LASCO and are likely related with the expansion of the prominence, being too diluted at these altitudes to produce significant \Lya\ emission. The subsequent images also show a radial intensity enhancement associated with the formation of the post-CME CS, and the subsequent trailing plasma blob moving at a projected speed of $\sim 210$\,\kms\ (as measured with the two UV frames acquired at 08:15 and 08:30\,UT). All these features (indicated with white arrows in Fig.~\ref{Fig:CMEdescription}) are also marked in different panels of Fig.~\ref{Fig:METISimages}.

\begin{figure*}
  \centering
  \includegraphics[width=0.75\textwidth]{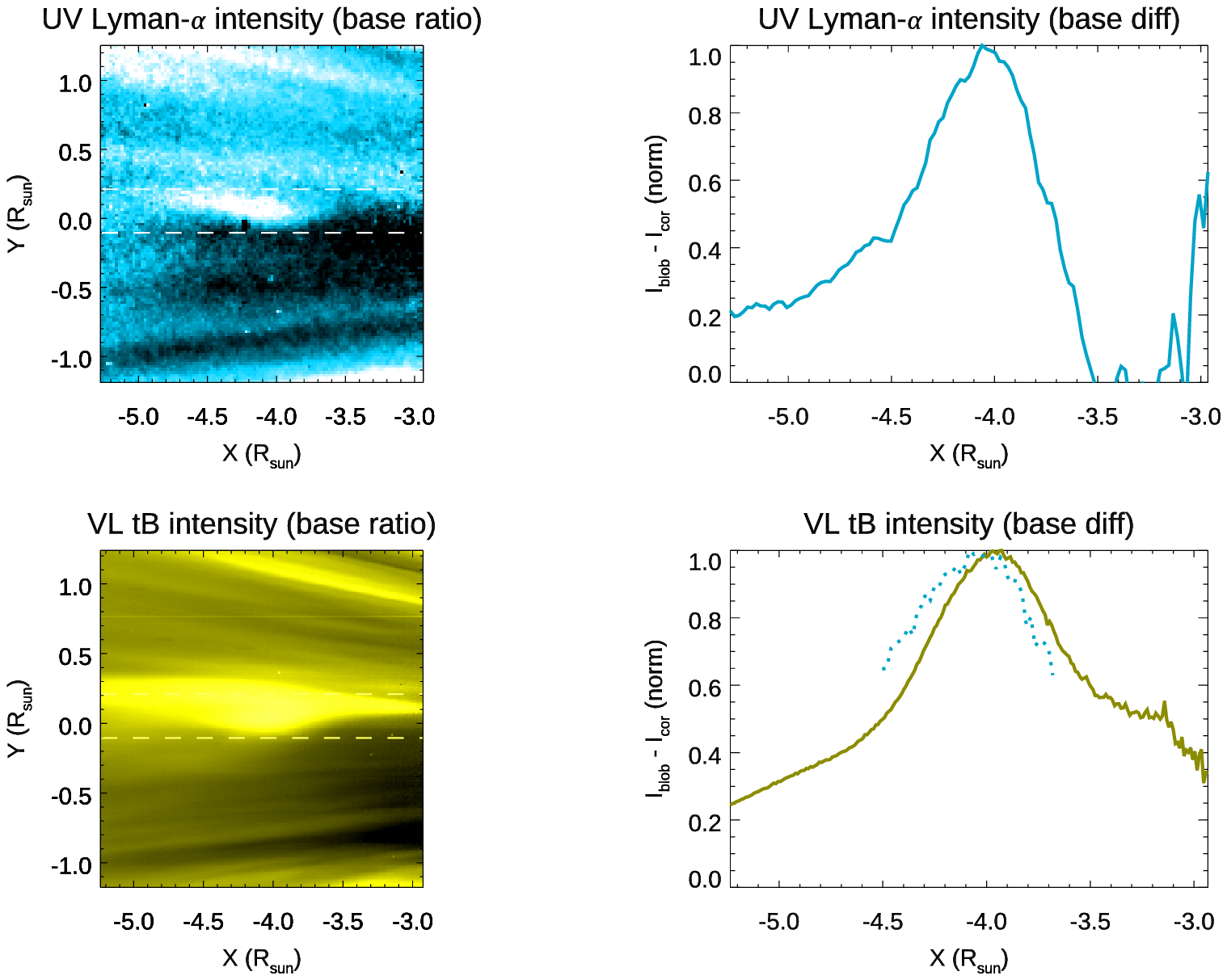}
  \caption{\textit{Left:} a zoom over the expanding small-scale post-CME plasma blob observed by Metis on February~13 between 08:15 and 08:46\,UT in the UV (top) and the VL (tB, bottom) channels (see also bottom right panel of Fig.~\ref{Fig:METISimages}). These images are obtained from the ratio between the actual frames and those acquired between 12:15 and 12:46\,UT on February~12; the horizontal dashed lines mark the area where the VL and UV intensities have been averaged to create the East-West profiles shown in the right panels. \textit{Right:} normalized UV (top) and VL (tB, bottom) East-West intensity profiles as obtained by averaging North-South over the areas shown with horizontal dashed lines in the left panels. These profiles are obtained after a subtraction between the actual frames and those acquired between 12:15 and 12:46\,UT on February~12. The dotted line in the bottom right panel shows a comparison to the square root of the UV profile shown in the top right panel (see text for explanations).}
  \label{Fig:Metisblob}
\end{figure*}
The presence of spatial gradients in the plane-of-sky (PoS) distribution of plasma velocities or temperatures should result in variations of the UV intensity distribution with respect to VL, related with the Doppler dimming of UV \Lya\ emission or the abundance of neutral H atoms. Nevertheless, a comparison between pB and UV images shows no evident differences between the appearance of CME features in the two channels that could suggest such gradients. On the other hand, the bottom left panels of Fig.~\ref{Fig:METISimages} show very interestingly that the post-CME radial feature that we identified as the CS appears as an intensity enhancement in the VL, and an intensity depletion in the UV \Lya. This difference is better shown in Fig~\ref{Fig:MetisCS}, showing a zoom over the coronal regions involved with the formation of this radial feature (left panels), and a vertical (North-South) cut proving the normalized intensity distributions across this feature in the VL (red lines) and in the UV \Lya\ (blue lines) before (top right) and after (bottom right) its formation. These observations are in very nice agreement with what previously observed with the UVCS instrument \cite[see e.g.][]{Lin_2005}. In particular, the depletion in the \Lya\ emission is a signature of higher radial velocities along the CS, and higher plasma temperatures, both related with on-going magnetic reconnection processes, and both resulting in a reduction of the observed \Lya\ emission.

In order to compare VL and UV observations of moving features, it is important to take into account that, considering the above CME and blob propagation speeds in the Metis FoV during the acquisition times, the observations are expected to be blurred by approximately 24--69 and 6--17 pixels in the VL and UV channels, respectively. These expected blurring in the VL channel are four times larger with respect to those in the UV, because the VL pixel resolution is two times better than UV, and the VL acquisition time is twice as long as the UV. Because the VL acquisition lasted $\sim 15$ minutes more than the UV acquisition, we expect displacements between the VL and UV images by $\sim 0.13$\,\rsun\ and $\sim 0.30$\,\rsun\ for the CME and the blob, respectively. We also remind that the effective Metis spatial resolution of the UV channel is degraded to 80\arcsec\,pix$^{-1}$ \citep{Antonucci-etal:2020}.

To reduce differences related with the different acquisition time intervals, for this sequence it is sufficient to average the two successive UV frames to cover with UV the same acquisition time covered by one full VL polarized observation, and avoid possible differences related with plasma motions. As an example, by averaging the two UV images acquired between 08:15--08:30\,UT and 08:31--08:46\,UT, it is possible to perform a comparison between the VL and UV normalized light distributions across the post-CME plasma blob, as it is shown in Fig.~\ref{Fig:Metisblob}, to be compared also with the bottom right panel of Fig.~\ref{Fig:METISimages}. The resulting radial distribution of the plasma blob emission appears to be broader in VL with respect to the UV image where the blob extension is instead more limited (Fig.~\ref{Fig:Metisblob}, right panels). In particular, the blob in the radial direction has an extension of 0.60\,\rsun\ in VL and 0.48\,\rsun\ in UV (by measuring the extension within 75\% of the intensity peak). Moreover, the blob in the UV appears to be radially shifted with respect to the VL by about 0.15\,\rsun. These interesting differences between the VL and UV light distributions can be explained as discussed below. 

A comparison between different VL and UV frames acquired during the blob propagation shows that its radial extension was not increasing significantly with time. In particular, from the analysis of the LASCO/C2 frames acquired on February~13 at 08:12\,UT and 08:48\,UT, it turns out that in this time interval the blob expanded radially by no more than $\sim 0.09$\,\rsun\ (corresponding to a projected expansion speed by $\sim 29$\,\kms), propagating radially by 0.42\,\rsun\ (corresponding to a projected propagation speed of 135\,\kms). Hence, the blob appeared to propagate with a very limited radial expansion rate ($\sim 20$\%), which means that the radial speed distribution across the blob can be considered constant. As a consequence, no significant radial variations of the \Lya\ emission across the blob are expected because of the Doppler dimming effect. Moreover, considering that the un-projected propagation speed of $\sim 210$\,\kms\ corresponds to a normalized \Lya\ Doppler dimming factor $\sim 0.3$ \citep[see e.g.][Fig. 3]{Capuano-2021}, and considering that the blob is brighter in the VL hence has also a larger plasma density, the observed \Lya\ intensity increase associated with the blob is expected to be mainly due to collisional rather than radiative excitation of H atoms, similarly to what has been previously discussed by \citet{Bemporad-etal:2018} for the \Lya\ observations of CME cores. Because the collisional component of the EUV-UV emission lines is roughly proportional to the square of the plasma density, while the VL emission is roughly proportional to the density, this could also explain why the radial extension of the UV blob appears much more limited with respect to what observed in VL.

To support this interpretation, the right panels of Fig.~\ref{Fig:Metisblob} show the normalized radial intensity distributions across the blob as obtained after subtraction of the pre-CME images for the UV (top) and VL (bottom, solid line) acquired between 12:15 and 12:46\,UT on February~12, as well as the square root of the UV distribution (bottom, dotted line). The square root of the normalized UV intensity profile appears much more similar to the VL intensity profile, supporting the interpretation that the UV emission is mostly collisional, but the evident shift by $\sim 0.15$\,\rsun\ between the two profiles requires a different explanation. This effect can be interpreted instead as an interesting signature of a temperature gradient across the blob. In particular, by assuming that the ascending top fraction of the plasma blob at 4.5\,\rsun\ has a typical coronal streamer temperature at this altitude by $1.0 \times 10^6$\,K \citep[see e.g.][]{Vasquez-2003}, a gradual temperature increase across the blob up to $\sim 1.55 \times 10^6$\,K at 3.9\,\rsun\ could explain the observed apparent shift between the UV and VL profiles, as determined with the H ionization equilibrium curve provided by the CHIANTI spectral code \citep{Dere-2019}. In fact, this temperature increase corresponds to a decrease in the density of neutral H atoms by about a factor 1.6: once the normalized VL and UV profiles in the bottom right panel of Fig.~\ref{Fig:Metisblob} are equalized at 4.5\,\rsun, this is the factor allowing to reproduce the UV to VL intensity jump at 3.9\,\rsun. We also notice that any possible radial variation of the \Lya\ Doppler dimming coefficient across the blob due to its radial expansion (neglected in the temperature estimate) would lead to a shift between the VL and UV emissions opposite to what observed (bottom right panel of Fig.~\ref{Fig:Metisblob}), by reducing more significantly the \Lya\ intensity in the blob regions located at higher altitudes and moving faster, and vice-versa. The possible physical explanation for this temperature gradient across the blob will be discussed in the last paragraph.

\subsection{SoloHI observations}
\label{sec:hi_obs}

\begin{figure}
  \centering
  \includegraphics[width=0.49\textwidth]{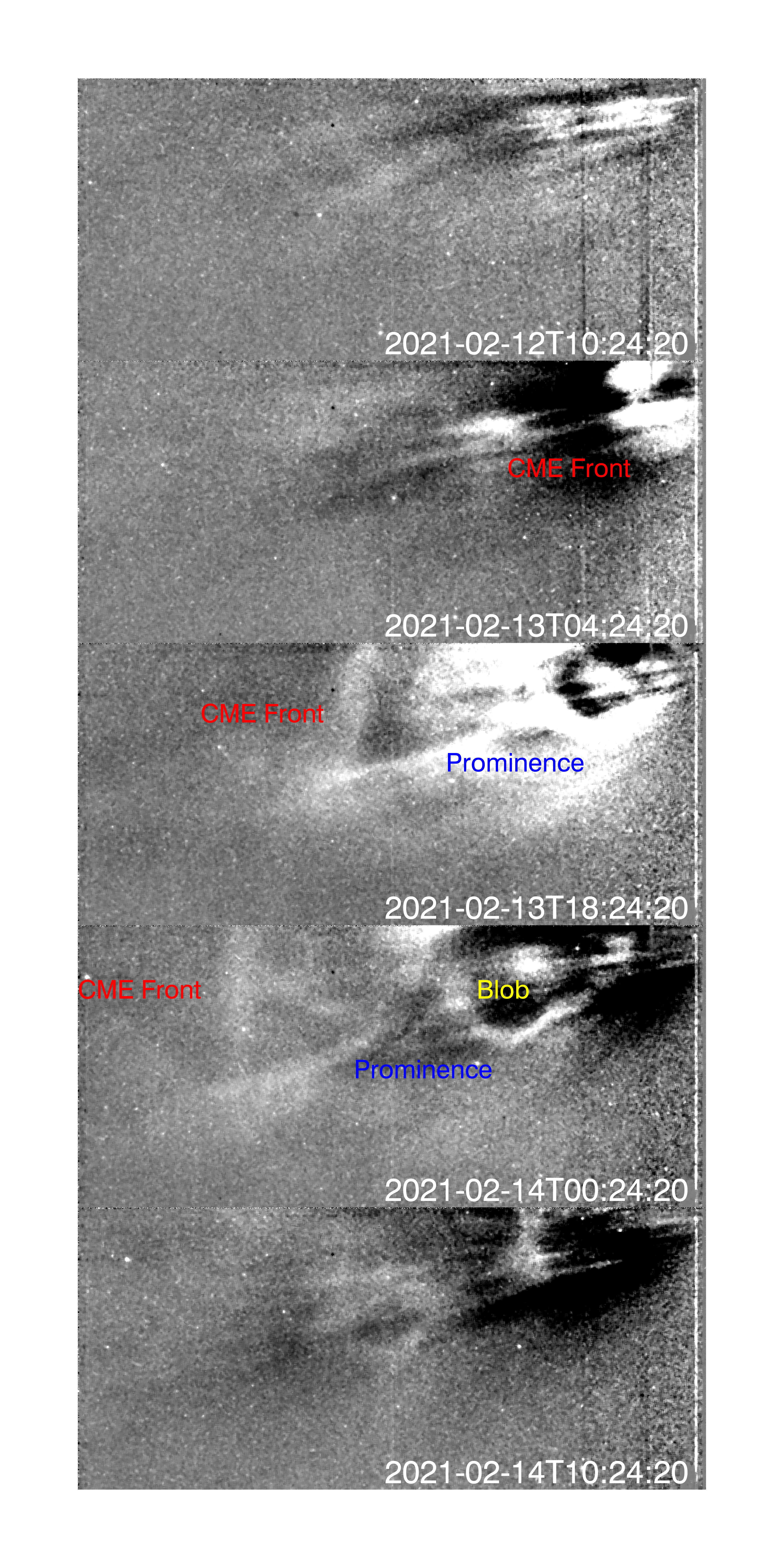}
  \caption{A series of SoloHI still frames showing the passing of the CME through Tile 1. Each frame is a cropped area of the full detector corresponding to all 2048 pixels in the x-direction and the top 920 detector pixels in the y. The CME front is labeled in red in the three frames where it is visible. The eruptive prominence (blue) and the following plasma blob (yellow), corresponding to structures in Metis (Fig.~\ref{Fig:METISimages}), are also shown.}
  \label{Fig:ShiStack}
\end{figure}

During the period of February~12-14, SoloHI was performing a test observing program with a 2 hour cadence on a single tile (Tile 1) of the detector. This is the bottom right tile in the overall mosaic, providing coverage of roughly 5\degr-25\degr\ in elongation and $-20\degr$-0\degr\ in latitude. This coverage was sufficient to observe much, but not all, of the CME front and associated material observed in Metis. In particular, Fig.~\ref{Fig:ShiStack} shows a series of snapshots of the SoloHI observations between February~12 and 14. The images have been processed to minimize the F-corona and highlight the small solar outflows visible in SoloHI before the CME appears. The front of the CME enters the instrument FoV just after 00:00\,UT on February~13. Complex structures are visible behind the leading edge, likely corresponding to both the CME flux rope and outflows associated with the restructuring behind the CME and shown in Fig.~\ref{Fig:METISimages}. 

\begin{figure}
  \centering
  \includegraphics[width=0.49\textwidth]{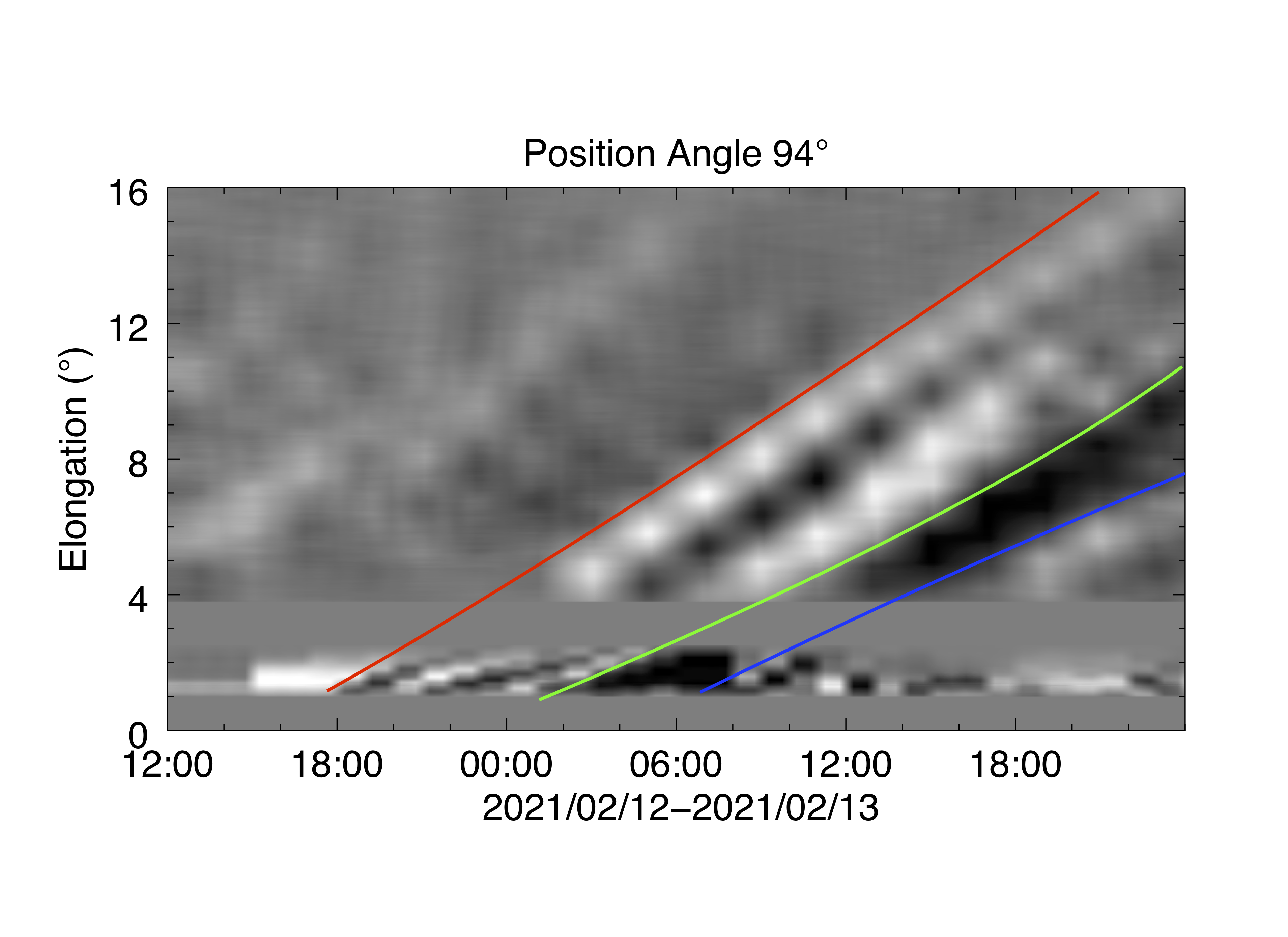}
  \caption{Combined Metis and SoloHI J-map at Position Angle 94\degr. The CME front (red), trailing edge (green), and the later eruptive prominence (blue) are all denoted with lines on the plot. The SoloHI FoV was cropped to 16\degr\ in elongation to allow the Metis FoV from $\sim 1.25\degr$ to 3\degr\ to be seen better.}
  \label{Fig:ComJmap}
\end{figure}

A distinct structure is visible around 10:00\,UT on February~13, although it does not become easily isolated until 18:24\,UT. This is believed to be related to the eruptive prominence seen in Metis at 06:30\,UT. Behind it, a small blob is also observed that is likely the SoloHI counterpart of the plasma blob highlighted in Fig.~\ref{Fig:METISimages} at 08:30\,UT in Metis.

To better illustrate the connection between structures seen in Metis and those observed in SoloHI, a J-map or height time plot featuring data from both instruments is shown in Fig.~\ref{Fig:ComJmap}, built by extracting intensity slices at position angle of 94\degr, or 4\degr\ below the equatorial plane. The map has a gap of approximately 2\degr\ in elongation between Metis and SoloHI, but even without direct overlap of each FoV it is possible to relate the features seen in each instrument in a manner consistent with a streamer blowout CME that has a gradual rise phase in the Metis FoV before being accelerated to a more constant speed in SoloHI. 
The CME front in Fig.~\ref{Fig:ComJmap} is denoted with a red line, and corresponds to a profile with an average speed of $\sim 250$\,\kms\ throughout the combined FoV. Everything between the red and green lines is considered to be either part of the CME flux rope or an associated outflow. The eruptive prominence is marked with the blue line and corresponds to a slower speed of $\sim 170$\,\kms.

\begin{table*}
\renewcommand{\arraystretch}{1.2}
\begin{center}
\caption{3D CME parameters as derived from the GCS fitting. 1st column: Date and time of the reconstruction in Metis. 2nd and 3rd columns: the longitude and latitude of the CME as observed from the Earth perspective. 4th column: tilt angle with respect with the solar equator. 5th column: height of the CME measured from the Sun center, in solar radii (\rsun). 6th and 7th columns: the aspect ratio and the half angle of the CME, respectively.}
\label{table: GCS}
\begin{tabular}{c c c c c c c}
\hline\hline
CME Date and time	&	Longitude	& Latitude	&	Tilt angle	&	Height & Ratio & Half angle	\\
(yyyy-mm-dd hh:mm)	&	(\degr)	   & (\degr)	&	(\degr)	        &	(\rsun)   &       & (\degr) \\
\hline	
2021-02-12 15:15	&	W\,60	& S\,06	&	0	&	4.042  &  0.4  &  30 	\\
2021-02-12 16:15	&	W\,60	& S\,06	&	0	&	4.328  &  0.4  &  30 	\\
2021-02-12 17:15	&	W\,60	& S\,06	&	0	&	4.642  &  0.4  &  30 	\\
\hline
\end{tabular}   
\end{center}
\end{table*}

\begin{figure*}
  \centering
  \includegraphics[width=0.3\textwidth]{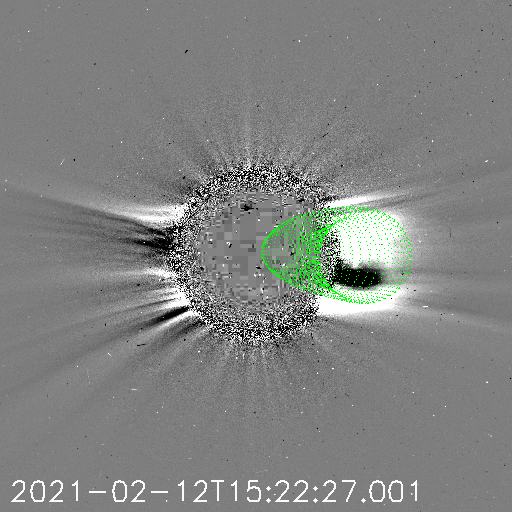}
  \includegraphics[width=0.3\textwidth]{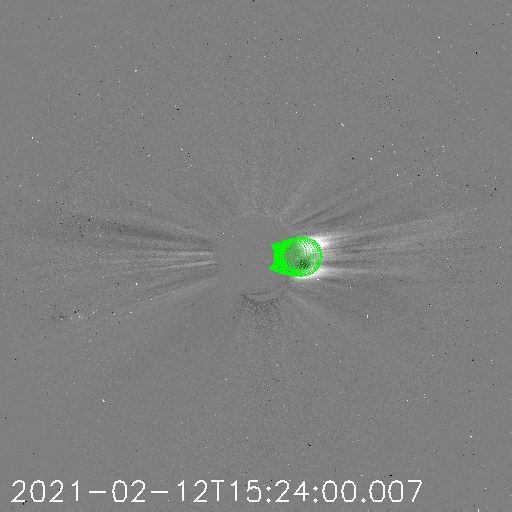}
  \includegraphics[width=0.3\textwidth]{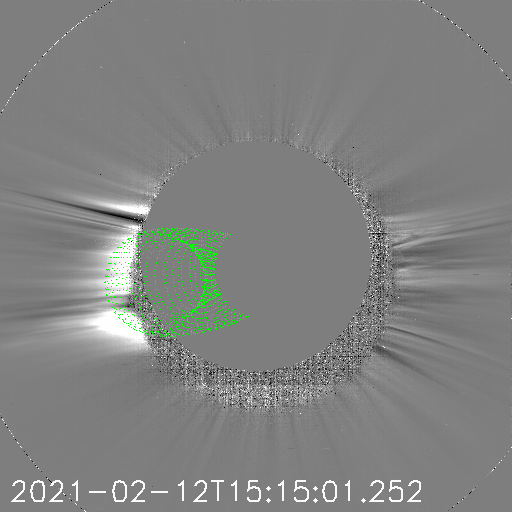}
  \includegraphics[width=0.3\textwidth]{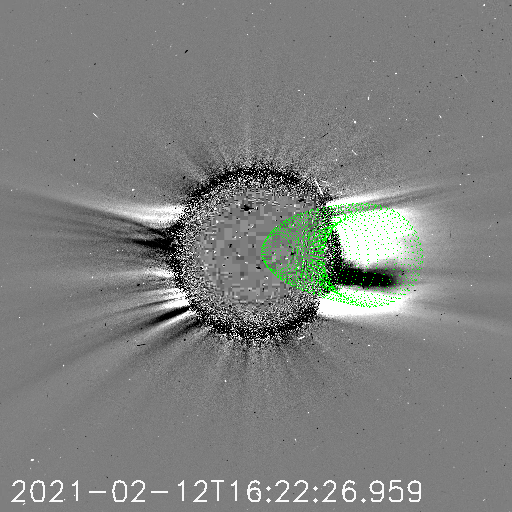}
  \includegraphics[width=0.3\textwidth]{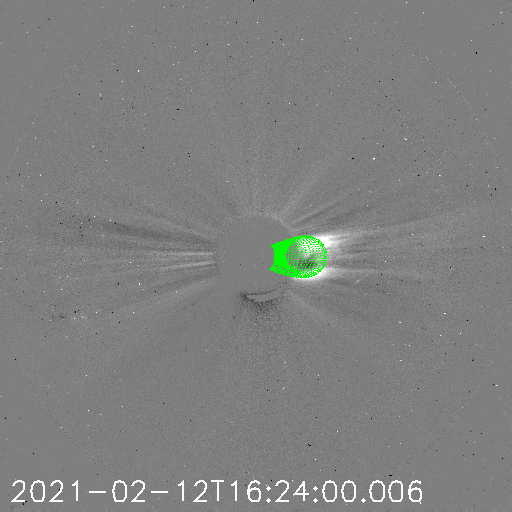}
  \includegraphics[width=0.3\textwidth]{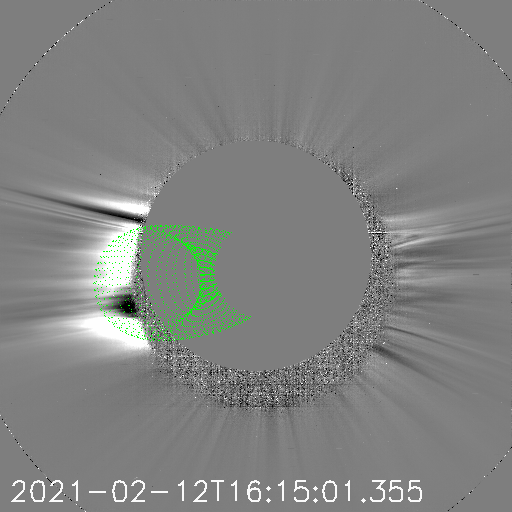}
  \includegraphics[width=0.3\textwidth]{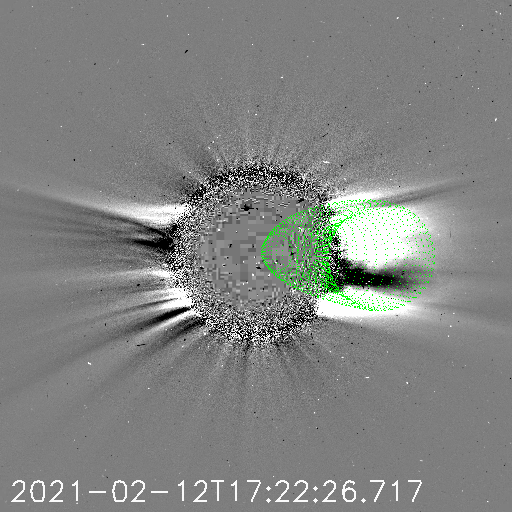}
  \includegraphics[width=0.3\textwidth]{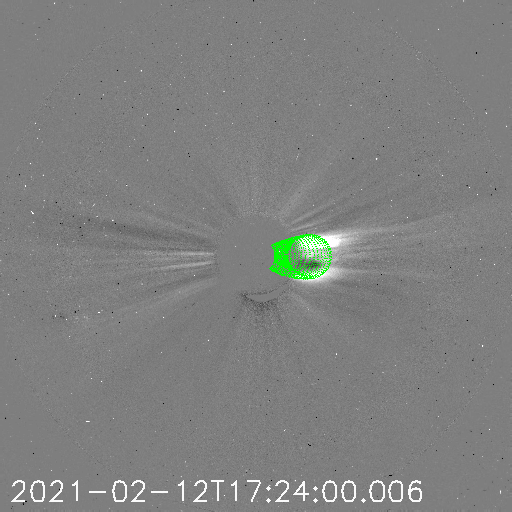}
  \includegraphics[width=0.3\textwidth]{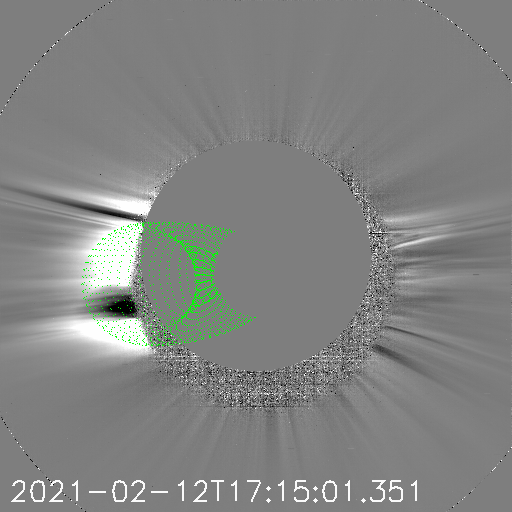}
  \caption{\textit{Upper row}: GCS reconstruction of the CME on February~12 observed by LASCO-C2 at 15:22 - 07:58\,UT (left panel), by COR2-A at 15:24 - 08:24\,UT (middle panel) and by Metis at 15:15 -12:15\,UT (right panel). \textit{Middle row}: GCS reconstruction of the CME on February~12 observed by LASCO-C2 at 16:22 - 07:58\,UT (left panel), by COR2-A at 16:24 - 08:24\,UT (middle panel) and by Metis at 16:15 -12:15\,UT (right panel). \textit{Lower row}: GCS reconstruction of the CME on February~12 observed by LASCO-C2 at 17:22 - 07:58\,UT (left panel), by COR2-A at 17:24 - 08:24\,UT (middle panel) and by Metis at 17:15 -12:15\,UT (right panel).}
  \label{Fig:GCSreconstr}
\end{figure*}

\section{Reconstructions with multiple observations}
\label{sec:3d_kinematics}

In order to identify the source regions of the above described events and their subsequent evolution, in this work we applied different reconstruction techniques to derive their 3D positions, namely: the graduated cylindrical shell (GCS) model \citep[see e.g.][]{Thernisien-2009, Thernisien-2011a} for the main CME, and triangulation \citep{Inhester-2006} for the erupting prominence.

\begin{figure*}
  \centering
  \includegraphics[width=0.7\textwidth]{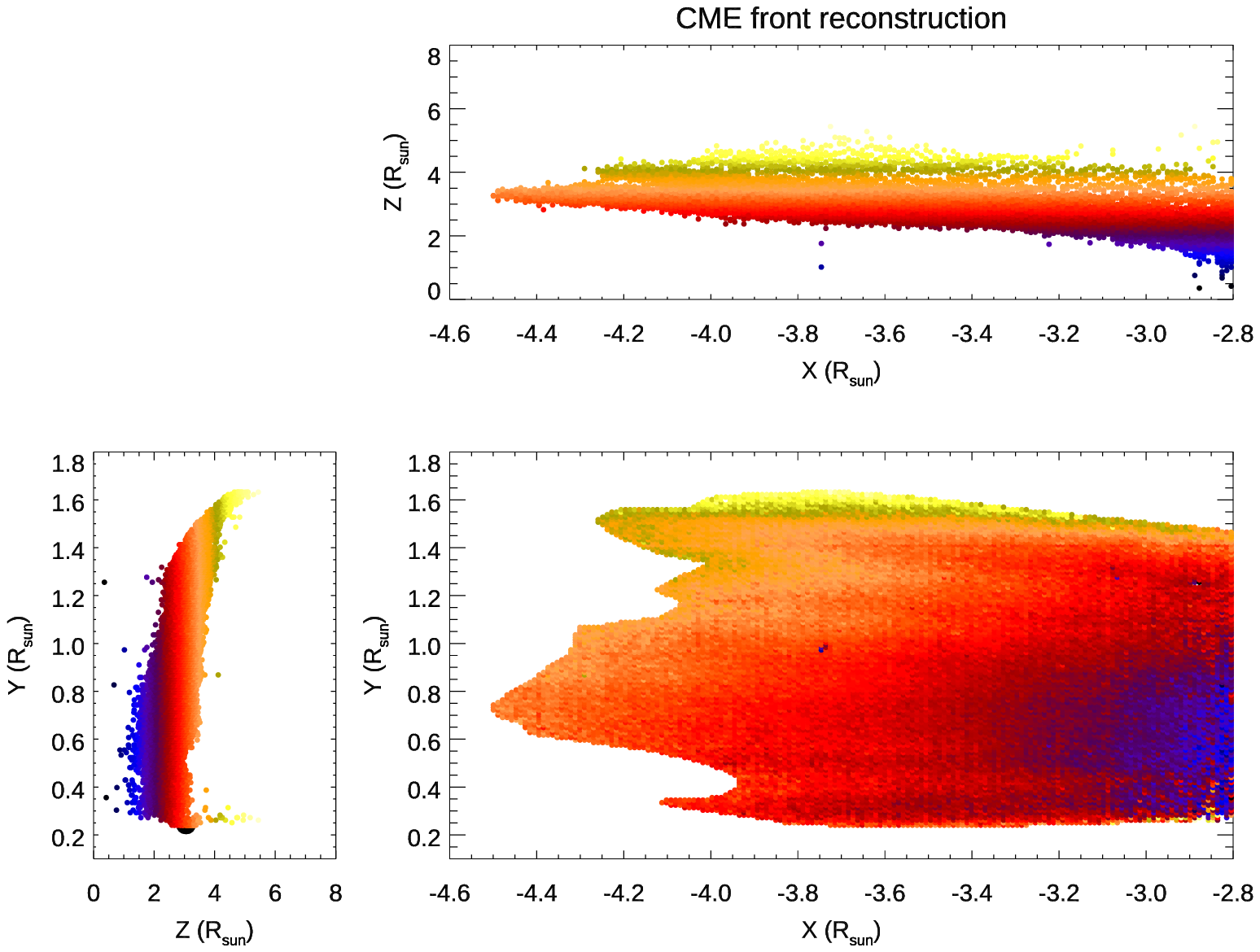}
  \includegraphics[width=0.7\textwidth]{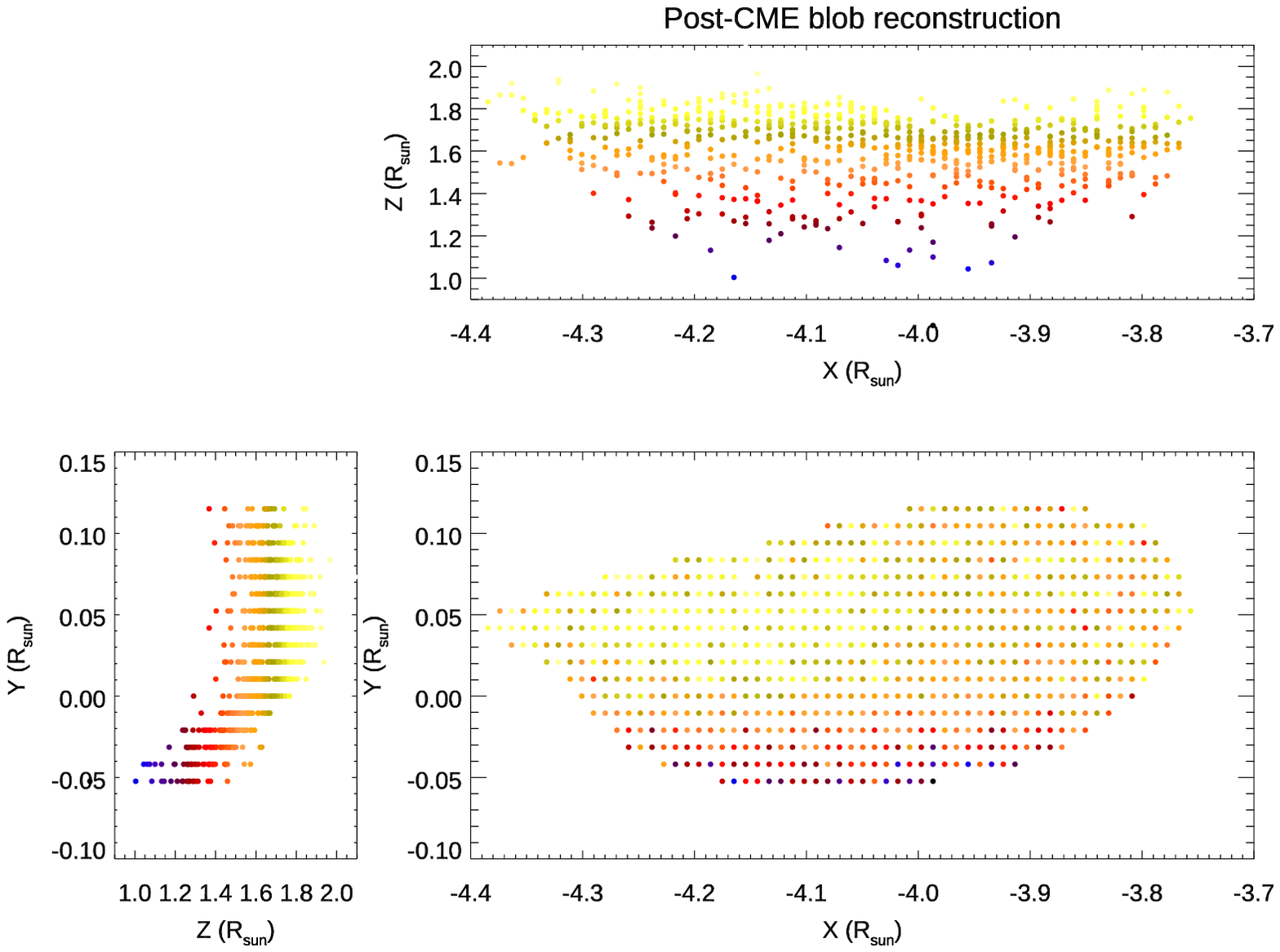}
  \caption{The 3D distribution of plasma emitting elements as obtained from the polarization ratio technique applied to the VL Metis observations of the CME front (top panels) and the post-CME blob (bottom panels). Different colors correspond to different distances from the PoS (located at $Z = 0$) as given in the left panels.}
  \label{Fig:polratio front}
\end{figure*}

The GCS model is meant to reproduce the large scale structure of flux rope-like CMEs. It consists of a tubular section forming the main body of the structure attached to two cones that correspond to the “legs” of the CME. Only the surface of the CME is modeled, and there is no rendering of its internal structure. This gives us information mostly on the propagation of the leading edge of the CME. The model fits the geometrical structure of CMEs as observed by different white-light coronagraphs. The output of the model are: the propagation longitude and latitude, half-angular width, aspect ratio, tilt angle with respect to the solar equator and the leading-edge height of the CME. The fitting can be done for multiple consecutive time steps in the corona giving the height/time profiles and how the CME geometrical and kinematic parameters evolve in the corona. 

We performed the GCS fitting for the CME observed by Metis at 15:15, 16:15, and 17:17\,UT (see Fig.~\ref{Fig:GCSreconstr}, right panels). The corresponding LASCO-C2 and COR2-A images are taken at 15:22, 16:22, 17:22\,UT (Fig.~\ref{Fig:GCSreconstr}, left panels) and 15:24, 16:24, 17:24\,UT (Fig.~\ref{Fig:GCSreconstr}, middle panels), respectively. As the light travels in around four minutes from SolO to Earth and STEREO-A, these images are taken 3 minutes (LASCO-C2) and 5 minutes (COR2-A) later than the corresponding Metis images. The results of the GCS reconstruction are provided in Table~\ref{table: GCS} in the Stonyhurst coordinate system. It is seen that the CME propagates on a radial direction along a longitude by 60\degr\ West and a latitude by 6\degr\ South. These coordinates are in good agreement with the location of the possible source region as observed by the PROBA2/SWAP and STEREO/EUVI instruments at approximately $\sim 20\degr-30\degr$ Eastward and $\sim 24\degr$ Southward (see Section \ref{sec:general}), indicating that in the early propagation phases the CME expanded more Westward and Northward. This non-radial expansion is also clearly suggested by the different locations of the EUV front as outlined in the sequence of STEREO/EUVI images (yellow lines in Fig.~\ref{Fig:EUVIfront}). Moreover, the height values provided in Table~\ref{table: GCS} correspond to an average radial speed of approximately 60\,\kms, even lower than previous estimates based on single images acquired by the LASCO and the Metis coronagraphs. These differences are likely related with the weakness of the CME front emission, leading to uncertainties in the position tracking that can be mitigated only by using multiple view-point observations. It is important to notice also that with such a low speed no significant Doppler dimming effect is expected in the \Lya\ emission observed by Metis, explaining the similarity between CME coronagraphic images acquired in the VL and UV channels (Fig.~\ref{Fig:METISimages}).

\begin{figure*}
  \centering
  \includegraphics[width=0.8\textwidth, bb = 30 5 500 235]{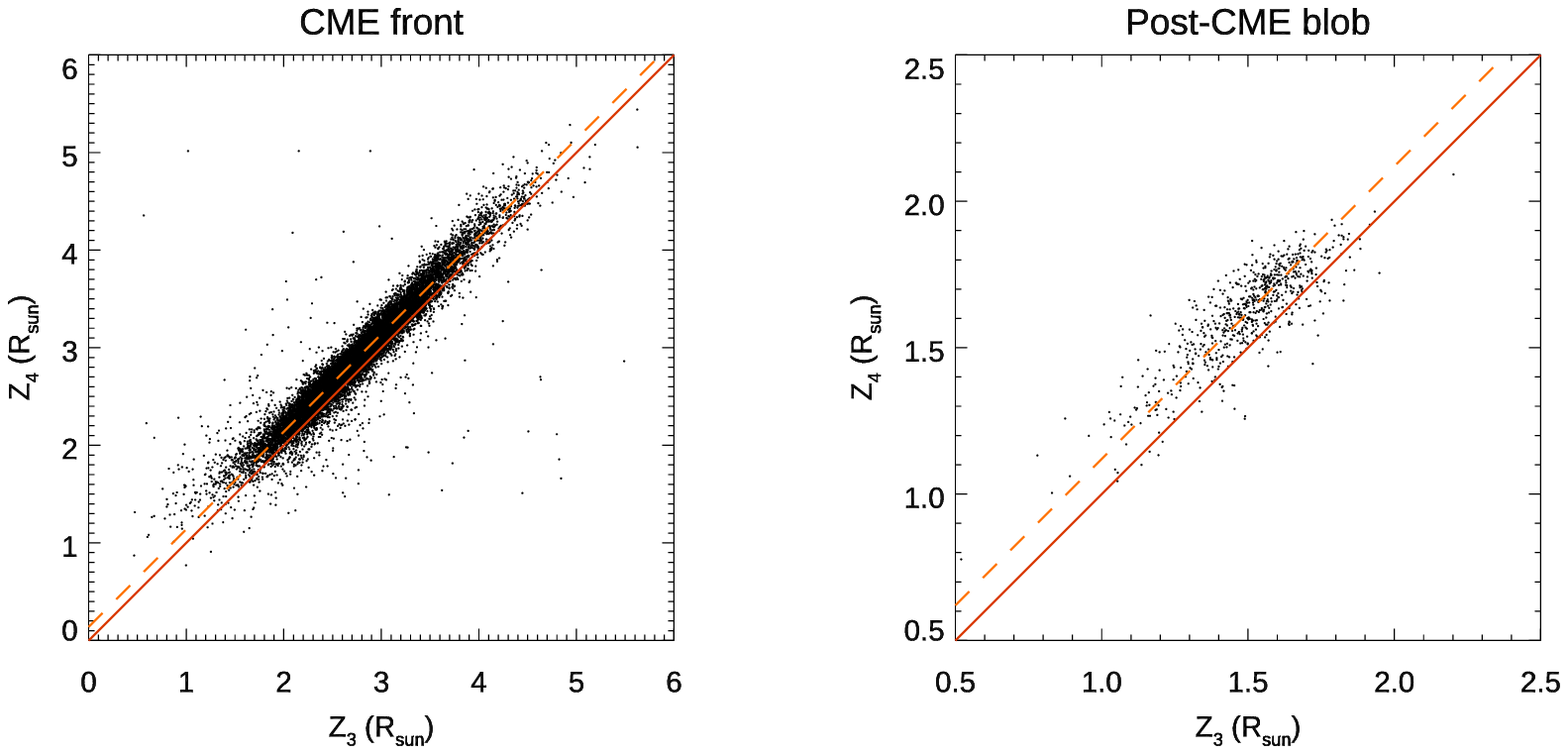}
  \caption{Scatter plot comparisons between the distributions of LoS positions as obtained with the polarization ratio technique applied to the same Metis data points by using four ($Z_4$) and three ($Z_3$) different orientations of the linear polarizer, for the CME front (left) and the post-CME blob (right). The solid lines show the reference curve for $Z_4 = Z_3$, while the dashed lines show the best fit.}
  \label{Fig:polratio blob}
\end{figure*}

In this work, we also performed triangulation on the leading edge of the erupting prominence observed by FSI \SI{17.4}{\nano\meter} at 19:15\,UT and by SWAP at 19:18\,UT. The triangulation method requires identification of the same point in the two images \citep[a process called tie-pointing, see e.g.][]{Inhester-2006}. Moreover, the method can be applied only for relatively smaller-scale and compact features (such as prominences or blobs), because a point-like geometry needs also to be assumed \cite{Liewer_2011}. This assumption is valid in this case, also considering that the separation angle between the SWAP and FSI LoS was small ($18^\circ$), and by tracking the outermost bright region of the prominence the errors are minimized. The 3D positions of LoS passing through the point that is visible in the two images are calculated, and the position of the intersection point in 3D space is determined. We performed the triangulation on the outermost bright point of the leading edge of the prominence observed by FSI and SWAP using the \textit{scc\_measure.pro} program of SolarSoft \citep{Thompson-2008, Thompson-2009}, which outputs the position of the feature in Stonyhurst coordinates \citep[longitude, latitude, and height from the Sun’s center, see][]{Thompson-2006}.

The images recorded at the time mentioned above showed best the erupting prominence in both FSI and SWAP field of view. Later on the leading edge of the eruption is hardly visible in SWAP images. The derived longitude, latitude and height are 85\degr\ West, 16\degr\ South and 1.447\,\rsun, respectively. This places the erupting prominence approximately 25\degr\ Westward and 10\degr\ Southward with respect to the main CME propagation direction as reconstructed by the VL coronagraphs. Considering also the value of the CME half angle of about 30\degr\ as provided by the GCS reconstruction, this suggests that the prominence erupted from a region that was located on the disk very close to the possible location of the flanks of the expanding CME, thus supporting the possibility that the final prominence destabilization was related with the nearby CME expansion.

\section{Reconstruction with polarization ratio}
\label{sec:metis_polariz}

In this work, the 3D reconstructions have been performed also with the polarization ratio technique first described by \citep{Moran-2004} and then applied by many different authors to reconstruct in 3D solar eruptions based on single viewpoint observations. This technique is particularly interesting for halo CMEs \citep[e.g.][]{Dolei-2014, Lu-2017}, and results from this technique were compared and validated with multiple viewpoint observations \citep[see e.g.][]{Mierla-2010, Mierla-2011, Feng-2013} and also with spectroscopic observations \citep{Susino-etal:2014}. Nevertheless, before Metis all the coronagraphic space-based data have been acquired so far in the VL and by using only three different angular orientations of the linear polarizer, typically separated by 60\degr\ or 120\degr. This number of exposures may result in errors in the estimate of the pB and polarization angle, as investigated with MHD numerical simulations by \citet{Pagano-2015} and more recently quantified by \citet{Inhester-2021}, errors that could be mitigated by acquiring a larger number of exposures.

Now, thanks to the Metis coronagraph, it is possible for the first time to have sequences VL of coronagraphic images acquired with four different orientations of a linear polarizer, separated by $\sim 45\degr$ \citep[see][for more details]{Antonucci-etal:2020}. Hence, for the first time the polarization ratio technique can be applied to pB and tB images derived from polarized sequences acquired with four different orientations of the linear polarizer. 
This is expected to provide a reduction of the uncertainties in the polarization measurement, hence smaller dispersion in the measurement of the LoS average position of the scattering plasma.

In order to quantify the advantages of using four polarimetric acquisitions instead of three, for the analysis performed here selected sequences of Metis VL polarized images have been combined in two different ways, considering the cases of both three and four polarizer orientations. 
Since Metis polarimetric observations are routinely performed using polarization angles $\delta_1=181.8\degr$, $\delta_2=49.1\degr$, $\delta_3=84.3\degr$, and $\delta_4=133.2\degr$, to simulate an acquisition obtained with only three polarizer orientations we neglected the fourth image of the selected sequences.
The Stokes vector images and, in turn, the pB and tB images were then derived from the three polarized images using the inverse of the theoretical M\"{u}ller matrix corresponding to the effective polarization angles $\delta_1$, $\delta_2$, and $\delta_3$.
Furthermore, in order to make a reliable comparison between the two cases, we also derived ``special'' pB and tB images from the full polarimetric sequence (i.e., including also the image corresponding to the $\delta_4$ orientation) by using the theoretical M\"{u}ller matrix instead of that measured during laboratory calibrations.
This alternative approach is necessary because the laboratory M\"{u}ller matrix has been derived from pixel-by-pixel measurements in order to account for inhomogeneities in the polarimetric response of Metis polarimeter, therefore it cannot be used to derive the Stokes images from a subset of the four polarized images of a nominal sequence.

Starting from the resulting pB and tB images as obtained with four different orientations of the linear polarizer, the polarization ratio technique has been applied to measure the LoS distribution of plasma emitting elements located in the CME front and in the the post-CME blob. In particular, for the CME front the technique has been applied to the polarized sequence acquired by Metis on February~12 between 20:15 and 20:45\,UT, after subtraction of the pre-CME sequence acquired on the same day between 13:15 and 13:45\,UT. On the other hand, for the post-CME blob the technique has been applied to the polarized sequence acquired by Metis on February~13 between 08:15 and 08:45\,UT, after subtraction of the previous sequence acquired on the same day between 07:15 and 07:45\,UT. Then for comparison, the reconstructions have been performed also by using the same images as obtained with only three different orientations of the linear polarizer.

The resulting 3D distributions are shown in Fig.~\ref{Fig:polratio front} for the CME front (top three panels) and the post-CME blob (bottom three panels). The Cartesian coordinates are the same as those employed in Fig.~\ref{Fig:Metisblob}. From these measurements, it turns out that the CME front plasma was located at an average heliocentric distance of $h_{\rm front} = 3.65$\,\rsun, and a distance from the PoS along the Metis LoS by $Z_{\rm front} = (2.87 \pm 0.57)$\,\rsun, while the post-CME blob was at an average heliocentric distance of $h_{\rm blob} = 4.04$\,\rsun, at an average distance from the PoS along the Metis LoS by $Z_{\rm blob} = (1.62 \pm 0.16)$\,\rsun. Hence, in the Stonyhurst coordinates, the CME front was expanding at an average longitude $\theta_{\rm front} = 68\degr$ West, while the post-CME blob was expanding at an average longitude $\theta_{\rm blob} = 85\degr$ West. These angles are in very good agreement with values provided by the 3D reconstruction of the CME with the GCS model, and with the location of the erupting prominence derived with triangulation. This means that the post-CME blob, whose origin is unclear being observed only in the coronagraphic images and not in the EUV disk images, was almost aligned with the previous propagation of the erupting prominence, thus suggesting a clear association between the two events.

It is also interesting to make a comparison between the LoS coordinate distributions $Z$ of the plasma emitting points as obtained by Metis by using images acquired with three ($Z_3$) or four ($Z_4$) different orientations of the linear polarizer. This interesting comparison, performed here for the very first time, is given by two scatter plots in Fig.~\ref{Fig:polratio blob}, for the CME front (left panel) and the post-CME blob (right panel). Each panel shows the ideal reference curve for $Z_4 = Z_3$ (solid lines), and the best linear fitting curves (dashed lines). From Fig.~\ref{Fig:polratio blob}, it turns out that on average the LoS measurements $Z_3$ as obtained only with three different orientations of the linear polarizer would be systematically underestimated by 0.14\,\rsun\ and 0.12\,\rsun\ respectively for the CME front and the post-CME blob. These underestimates correspond to relative errors in the LoS positioning by 4.9\% and 7.4\%, respectively. This quantifies the improvement in the 3D reconstructions of CMEs based on the polarization ratio technique applied to the Metis data.

\section{Summary \& conclusions}
\label{sec:conclusions}

This paper analyzes a complex sequence of three successive eruptive events that occurred on the Sun between 2021 February~12--13. During these days, the Sun released a slow and accelerating Coronal Mass Ejection (CME), followed by a nearby prominence eruption and a trailing plasma blob. The events have been observed not only by ground- and space-based instruments located along the Sun-Earth line, but also by STEREO-A and the SolO spacecraft, located at separation angles with the Earth by 55.86\degr\ East and 161.6\degr\ East, respectively. The analysis presented here focused mostly on the data acquired by remote sensing instruments on-board SolO, supported also by data acquired by other instruments. The main results can be summarized as follows:\\
\begin{itemize}
    \item combination of images acquired by different view points with triangulation (applied to FSI data) and GCS fitting technique (applied to Metis data) allowed us to reconstruct the 3D trajectory and source regions of the CME and the erupting prominence, while the 3D location of the plasma blob was inferred with the polarization ratio technique; the CME turns out to originate from the Western hemisphere (as seen from the Earth) and then propagating Westward and Northward, the following prominence erupts closer to the West limb (as seen from the Earth) propagating Northward, the subsequent blob follows approximately the same trajectory as the prominence;\\
    \item these results once again confirm the importance of 3D reconstructions to measure the real propagation speed and direction of solar eruptions, that were both different from what determined from single view-point observations; this is important not only for Space Weather applications, but also for physical interpretation of the observed events: in this case the Westward early CME propagation and its angular extension support a possible cause-effect association with the subsequent destabilization of the erupting prominence, making these two candidate sympathetic events; moreover, the coalignment between the prominence and the following post-CME blob also suggests that the blob originating in the corona is likely due to magnetic reconnection occurring in the corona after the transit of the prominence;\\
    \item all the events have been also observed by the SoloHI instrument, and despite the existing gap between the Metis and SoloHI instrument field-of-views, the combined J-maps allow us to clearly track the same features in the two instruments;\\
    \item for the first time a post-CME Current Sheet (CS) has been observed in the intermediate corona with a multi-channel coronagraph in the VL and UV \Lya; this radial feature appears as a classical intensity enhancement in the VL (due to the plasma compression going on in the CS), and an intensity depletion in the UV \Lya\ (due to higher radial speeds and temperatures of plasma propagating along the CS), thus confirming previous UVCS observations of similar features in the H {\sc I} Ly-$\alpha$ \cite[][]{Lin_2005} and also in the Ly-$\beta$ \cite[][]{Bemporad-etal:2006} lines;\\
    \item for the first time the images acquired by the two Metis channels in VL and UV have been combined to measure the plasma temperature gradient across the post-CME blob; the observed shift between the VL and UV blob location was explained here by assuming that, because of Doppler dimming, the \Lya\ emission is entirely collisional; with this assumption (and by also assuming ionization equilibrium) the shift is explained by a radial temperature increase across the blob from $1.0 \times 10^6$\,K in the ascending top part up to $\sim 1.55 \times 10^6$\,K in the lower bottom part;\\
    \item the polarization ratio technique applied to Metis images provides values for the 3D location of the CME that are in very nice agreement with 3D reconstructions obtained with other methods; moreover, this work quantifies for the first time the improvement in the 3D reconstruction capabilities based on images acquired with four (instead of three) different polarization angles.
\end{itemize}
Before concluding, it is interesting to discuss a possible explanation and the implications for the measured temperature gradient across the post-CME blob. Similar small-scale post-CME blobs are usually interpreted as a result of post-CME coronal rearrangement and magnetic reconnection \citep[e.g.][]{Lynch-2013}, in particular along the newly formed post-CME current sheet \citep[see e.g.][]{Liu2009, Chae2017}. These blobs are typically hardly observed in EUV disk images \citep{Schanche2016}, making it difficult to understand their origin, and are typically observed in post-CME CS \citep[see e.g.][]{Ko_2003, Vrsnak_2009}. Recently, \citet{Lee2020} reported the identification of the same plasma-blobs observed by LASCO also in the ground-based images acquired by the K-Cor coronagraph, demonstrating that their formation is likely due to tearing instability \citep{Shibata2001, Barta_2008} occurring just above the tip of post-CME CS, leading to magnetic reconnection accelerating in most cases these blobs upward. This is also in agreement with \citet{Reeves-2019} who studied the formation of dense plasmoid-like structures breaking-up the post-CME current-sheet by tearing instability, whose plasma is heated predominantly by adiabatic compression rather than ohmic heating, thus implying that the onset of the plasmoid instability increases the heating of the surrounding plasma by adiabatic compression as the inflow velocity increases.

These previous findings suggest that the temperature gradient reported here for the first time could be the result of post-eruption magnetic reconnection driven by tearing instability and occurring in the corona just below the plasma blob. This reconnection is thus responsible for the outward acceleration of the blob and also for additional heating of the plasma by adiabatic compression, resulting in the observed temperature gradient. Future observations of similar blobs by Metis will be helpful to confirm or refute this scenario.

\begin{acknowledgements}
  The authors acknowledge important suggestions from the anonymous Referee in the identification of different features observed in the coronagraphic images. S.J. acknowledges the support from the Slovenian Research Agency No. P1-0188. Solar Orbiter is a space mission of international collaboration between ESA and NASA, operated by ESA.  The Metis programme is supported by the Italian Space Agency (ASI) under the contracts to the co-financing National Institute of Astrophysics (INAF): Accordi ASI-INAF N. I-043-10-0 and Addendum N. I-013-12-0/1, Accordo ASI-INAF N.2018-30-HH.0  and under the contracts to the industrial partners OHB Italia SpA, Thales Alenia Space Italia SpA and ALTEC: ASI-TASI N. I-037-11-0 and ASI-ATI N. 2013-057-I.0. Metis was built with hardware contributions from Germany (Bundesministerium f\"{u}r Wirtschaft und Energie (BMWi) through the Deutsches Zentrum f\"{u}r Luft- und Raumfahrt e.V. (DLR)), from the Academy of Sciences of the Czech Republic (Czech PRODEX) and from ESA.
  
  The EUI instrument was built by CSL, IAS, MPS, MSSL/UCL, PMOD/WRC, ROB, LCF/IO with funding from the Belgian Federal Science Policy Office; the Centre National d’Etudes Spatiales (CNES); the UK Space Agency (UKSA); the Bundesministerium f\"{u}r Wirtschaft und Energie (BMWi) through the Deutsches Zentrum f\"{u}r Luft- und Raumfahrt (DLR); and the Swiss Space Office (SSO). The ROB team thanks the Belgian Federal Science Policy Office (BELSPO) for the provision of financial support in the framework of the PRODEX Programme of the European Space Agency (ESA) under contract numbers  4000134474, 4000134088, and 4000136424.
  
  The SoloHI instrument was designed and built at the U.S. Naval Research Laboratory and supported by NASA under contract NNG09EK11I. R.C. and P.H are supported by SoloHI.
  
  The SOHO/LASCO data used here are produced by a consortium of the Naval Research Laboratory (USA), Max-Planck-Institut f\"{u}r Aeronomie (Germany), Laboratoire d'Astronomie (France), and the University of Birmingham (UK)\footnote{Please note that MPI for Aeronomy (MPAe) has changed its name to the Max-Planck-Institut for Sonnensystemforschung (MPS) and that the Laboratoire d'Astronomie (LAS) has changed its name to the Laboratoire d'Astrophysique Marseille (LAM)}. SOHO is a project of international cooperation between ESA and NASA.
  
  The SECCHI data are produced by an international consortium of the NRL, LMSAL, and NASA GSFC (USA), RAL and U. Bham (UK), MPS (Germany), CSL (Belgium), IOTA, and IAS (France).
  
  PROBA2/SWAP is a project of the Centre Spatial de Liege and the Royal Observatory of Belgium funded by the Belgian Federal Science Policy Office (BELSPO).


Courtesy of the Mauna Loa Solar Observatory, operated by the High Altitude Observatory, as part of the National Center for Atmospheric Research (NCAR). NCAR is supported by the National Science Foundation.

This work utilizes GONG data from NSO, which is operated by AURA under a cooperative agreement with NSF and with additional financial support from NOAA, NASA, and USAF.

J.K. acknowledges the project VEGA 2/0048/20.
\end{acknowledgements}


\bibliographystyle{aa}

\bibliography{metis_second_cme.bib}

\end{document}